\documentclass[aps,prb,reprint,groupedaddress,noeprint]{revtex4-2}
\usepackage{amsmath,amssymb}
\usepackage{graphicx}
\usepackage[colorlinks=true]{hyperref}

\providecommand{\abs}[1]{\lvert#1\rvert}

\begin{document}

\title{Field evolution of magnetic phases and spin dynamics in the honeycomb lattice magnet Na$_2$Co$_2$TeO$_6$: $^{23}$Na NMR study}

\author{Jun~Kikuchi}
\email[]{jkiku@meiji.ac.jp}
\author{Takayuki~Kamoda}
\author{Nobuyoshi~Mera}
\author{Yodai~Takahashi}
\author{Kouji~Okumura}
\author{Yukio~Yasui}
\affiliation{Department of Physics, School of Science and Technology, Meiji University, Kawasaki 214-8571, Japan}

\date{\today}

\begin{abstract}
We report on the results of $^{23}$Na NMR in the honeycomb lattice magnet Na$_2$Co$_2$TeO$_6$ which has been nominated as a Kitaev material. 
Measurements of magnetic shift and width of the NMR line as functions of temperature and magnetic field show that a spin-disordered phase does not appear up to a field of 9~T. 
In the antiferromagnetic phase just below the N\'eel temperature $T_N$, we find a temperature region extending down to $\sim T_N/2$ where the nuclear spin-lattice relaxation rate $1/T_1$ remains enhanced and is further increased by magnetic fields.  
This region crosses over to a low temperature region characterized by the rapidly decreasing $1/T_1$ which is less field-sensitive. 
These observations suggest incoherent spin excitations with a large spectral weight at low energies in the intermediate temperature region transforming to more conventional spin-wave excitations at low temperatures. 
The drastic change of the low-energy spin dynamics is likely caused by strong damping of spin waves activated only in the intermediate temperature region, which may be realized for triple-$\mathbf{q}$ magnetic order possessing partially-disordered moments as scattering centers of spin waves. 
In the paramagnetic phase near $T_N$, dramatic field suppression of $1/T_1$ is observed. 
From analysis of the temperature dependence of $1/T_1$ based on the renormalized-classical description of a two-dimensional quantum antiferromagnet, we find the field-dependent spin stiffness constant that scales with $T_N$ as a function of magnetic field. 
This implies field suppression of the energy scale characterizing both two-dimensional spin correlations and three-dimensional long-range order, which may be associated with an increasing effect of frustration in magnetic fields. 
\end{abstract}

\maketitle

\section{Introduction}

The search for novel quantum phases in frustrated magnets has long been the subject of intense studies since the first proposal of a resonating valence bond state in a triangular lattice antiferromagnet \cite{Anderson_RVB}. 
One of the most intriguing phases to be sought is a quantum spin liquid (QSL) which breaks no spontaneous symmetry and is characterized by topological quantities. 
Since QSLs are expected to appear under the influence of strong frustration and quantum fluctuations both suppressing magnetic long-range order (LRO), frustrated quantum-spin systems on low-dimensional lattices have been studied extensively from experimental and theoretical viewpoints \cite{Lacroix_Frustration,Balents_SL,Savary_QSL,Broholm_QSL}. 

A honeycomb lattice has the smallest coordination number among two-dimensional (2D) lattices and hence strong quantum effects are expected. Despite such fundamental importance, honeycomb lattice magnets seem less explored compared with triangular and square lattice magnets.
The honeycomb lattice is bipartite and is not frustrated for nearest-neighbor interactions like the square lattice, but the presence of further neighbor interactions introduces frustration, leading to various competing phases in both classical and quantum cases \cite{Rastelli79,Fouet01,Mulder10,Albuquerque11,Reuther11,Li12}. 
In the frustrated spin-1/2 Heisenberg model, quantum fluctuations are sufficient to destroy magnetic LRO in some parameter regions and gives rise to disordered ground states such as a gapped QSL and a dimer or plaquette valence-bond solid. 
In the case of $XY$ anisotropy, the honeycomb lattice magnets provide a playground for studying the fascinating Berezinskii-Kosterlitz-Thouless transition driven by topological excitations. 

Another direction of research on the honeycomb lattice magnets has been prompted by an exactly solvable quantum spin-1/2 model formulated by Kitaev \cite{Kitaev}. 
The model consists of nearest-neighbor bond-dependent Ising interactions of which easy axes are mutually orthogonal and is strongly frustrated. 
It has a gapped QSL ground state with fractionalized Majorana fermion excitations coupled to a $Z_2$ gauge field and has attracted growing interest in recent years \cite{Winter_Kitaev,Takagi_Kitaev,Janssen19,Motome_Kitaev,Trebst_Kitaev}. 

Materials realizations of the Kitaev model have been proposed for $4d$ and $5d$ transition-metal compounds including ions with a low-spin $d^5\,(t_{2g}^5)$ configuration \cite{Jackeli09,Chaloupka10}. 
These ions can host spin-orbital entangled pseudospin $j_\mathrm{eff}=1/2$ in an octahedral crystal field \cite{AB}, which enables bond-dependent coupling via anisotropic electronic wave functions. 
Extensive researches on the candidate materials such as Na$_2$IrO$_3$ and $\alpha$-RuCl$_3$ revealed that most of them display magnetic LRO, indicating the importance of non-Kitaev interactions in real materials. 
Generalized Kitaev models including the Heisenberg and off-diagonal exchange terms were developed to show that magnetic LRO is largely stabilized by these additional interactions but the QSL survives in small but finite regions of the parameter space \cite{Chaloupka13,Rau14}. 

In spite of the absence of a QSL phase, the relevance of Kitaev physics to magnetic properties of the candidate materials is recognized in many respects. 
One such example is an unconventional continuum of the magnetic excitations in $\alpha$-RuCl$_3$ and Na$_2$IrO$_3$ reminiscent of itinerant Majorana-fermion bands expected in the Kitaev QSLs \cite{Banerjee17,Banerjee18,KimJ20}. 
The two-step release of magnetic entropy is also interpreted as a signature of fractionalization of spin degrees of freedom \cite{Kubota15,Mehlawat17}. 
The most prominent hallmark would be the field-induced disordered phase of $\alpha$-RuCl$_3$ found above the in-plane critical field of 7$-$8~T \cite{Johnson15,Sears17,Wolter17}. 
Because of a close connection of this phase to field-driven QSL states in the Kitaev model \cite{Kitaev,Hickey19}
as well as the robustness of the excitation continuum against a field \cite{Banerjee18}, 
$\alpha$-RuCl$_3$ has been considered proximate to the Kitaev QSL. 

Recently, $3d$ transition-metal compounds including Co$^{2+}$ ions with a high-spin $d^7\,(t_{2g}^5 e_{g}^2)$ configuration have been suggested to be more suitable for the Kitaev magnet \cite{Liu18,Sano18,Liu20,Liu21}. 
The compound Na$_2$Co$_2$TeO$_6$ has received special attention among the Co candidates because 
it shows a field-induced transition from an antiferromagnetic to a putative spin-disordered phase similar to $\alpha$-RuCl$_3$. 
The crystal structure of Na$_2$Co$_2$TeO$_6$ belongs to the hexagonal space group $P6_322$ (No.~182) \cite{Viciu07,Lefrancois16,Bera17,Xiao19,Samarakoon21}. The CoO$_6$ octahedra comprised of two independent Co1 and Co2 sites share their edges to form nearly ideal honeycomb lattices in the $ab$ planes. 
Stacking of the honeycomb layers in the $c$ direction is such that the six-fold screw axis going through the Co1 site transforms a honeycomb lattice to the next layer. The Co2 site is hence on top of the Te site which is at the center of the adjacent Co hexagon. Na sites are in between the honeycomb layers and are partially occupied. 
Structural analysis indicated that a stacking fault and intermixing of cations in the honeycomb layers are not evident \cite{Bera17,Viciu07}. 

\begin{figure}
\includegraphics[width=6.5cm]{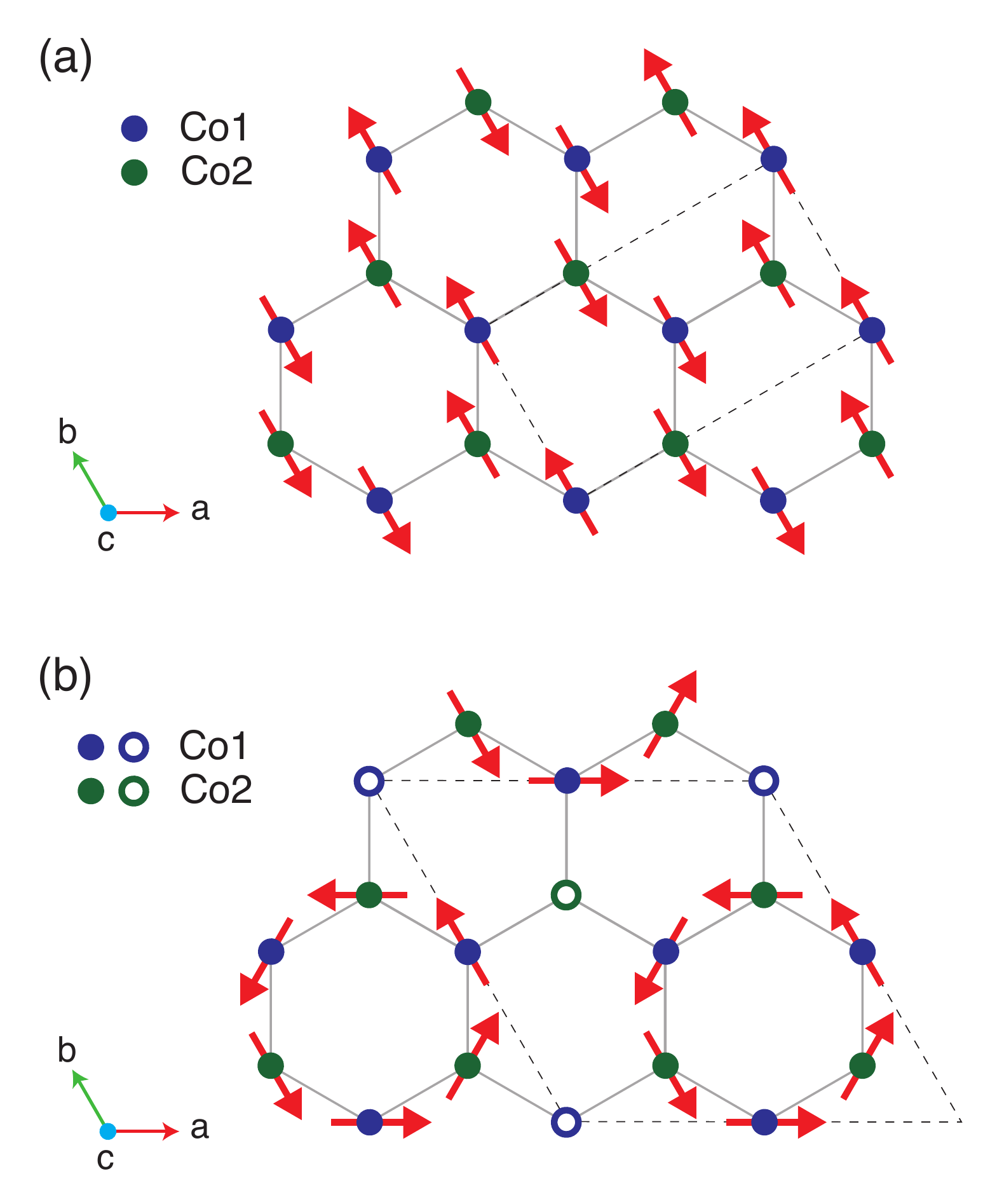}
\caption{\label{fig:ZigzagTripleQ}Magnetic structures proposed for Na$_2$Co$_2$TeO$_6$. The arrows indicate in-plane ordered moments. (a) Zigzag magnetic structure described by a propagation vector $\mathbf{Q}=(\frac{1}{2},0,0)$.  The dashed line represents the magnetic unit cell with a two-fold screw axis normal to the $ab$ plane at the origin. (b) An example of the triple-$\mathbf{q}$ magnetic structure. The dashed line represents the magnetic unit cell with a six-fold screw axis $6_3$ at the origin. Spinfull (``spinless'') Co atoms are shown by solid (open) circles. Note that only three quarters of Co atoms have an in-plane moment. 
Each Co atom including spinless one may have a $c$-axis N\'eel component allowed by $C_3$ symmetry.}
\end{figure}

The Co$^{2+}$ ions host pseudospin $1/2$ as revealed from the observation of spin-orbit excitations at 22$-$23~meV via inelastic neutron scattering (INS) \cite{Songvilay20,KimC22}. 
Antiferromagnetic (AFM) LRO characterized by a propagation vector $\mathbf{Q}=(\frac{1}{2},0,0)$ is observed below the N\'eel temperature $T_N\approx 27$~K at zero field \cite{Lefrancois16,Bera17,Samarakoon21}. 
The proposed magnetic structures are shown schematically in Fig.~\ref{fig:ZigzagTripleQ}.
The single-$\mathbf{q}$ collinear structure [Fig.~\ref{fig:ZigzagTripleQ}(a)] has a zigzag spin arrangement similar to Na$_2$IrO$_3$ \cite{Ye12} and $\alpha$-RuCl$_3$ \cite{Johnson15}; ferromagnetic zigzag chains running along the $b$ axis (perpendicular to $\mathbf{Q}$) align alternately in the direction of $\mathbf{Q}$ (parallel to $a^*$) with the ordered moment either pointing along the $b$ axis \cite{Lefrancois16,Bera17} or lying in the $bc$ plane \cite{Samarakoon21}. 
The triple-$\mathbf{q}$ noncollinear structure [Fig.~\ref{fig:ZigzagTripleQ}(b)] is formed by superposing the zigzag structure with $\mathbf{Q}=(\frac{1}{2},0,0)$ and the equivalents related by $C_3$ rotation \cite{Chen21}. 
The ordered moment on the Co1 site is larger by a factor of 1.1$-$1.2 than that on the Co2 site, 
reflecting the fact that the ordered phase is in a strict sense ferrimagnetic \cite{Yao20}. 

The magnetic susceptibility shows easy-plane anisotropy nominally described by a direction dependent Weiss temperature \cite{Xiao19,Yao20} suggesting anisotropic exchange interactions. 
Standard analysis in evaluating exchange parameters from the Curie-Weiss fit of the susceptibility was unsuccessful owing to the presence of competing interactions and an effect of low-lying spin-orbit excited states \cite{Liu20,Liu21}.
Powder INS experiments have been attempted to extract a set of parameters which consistently describes the magnetic excitation spectrum based on the linear spin-wave theory \cite{Songvilay20,Lin21,Samarakoon21,KimC22,Sanders22}. 
The generalized Kitaev model rather than the frustrated Heisenberg model is preferred, but the results are still diverse; even the sign of the leading interaction is unsettled, although theories predict a dominant ferromagnetic Kitaev interaction \cite{Liu18,Sano18}. 
A recent INS study on single crystals has revealed unusual features of the magnetic excitations: the existence of an incoherent continuum around $\mathbf{Q}=(\frac{1}{2},0,0)$ persisting down to $\sim T_N/2$ and the formation of a spin-wave mode below that temperature \cite{Chen21}. 

One of the most attracting features of Na$_2$Co$_2$TeO$_6$ is an anisotropic field response of the AFM phase resembling $\alpha$-RuCl$_3$. The AFM order of Na$_2$Co$_2$TeO$_6$ is suppressed by a moderate in-plane field, above which a QSL state is expected to emerge \cite{Yao20,Hong21,Lin21}. 
Yao and Li are the first who reported strong suppression of AFM order by an in-plane field and related anomalies in magnetization as well as in magnetic specific heat and suggested the existence of a high-field spin-disordered phase \cite{Yao20}. They also found that the magnetization jump detected in powders around 6~T \cite{Lefrancois16,Bera17} is observable only for the field $\mathbf{B}\parallel a^*$, i.e., perpendicular to the zigzag-ordered moment, which indicates that the transition is not of spin-flop type. 
Subsequent magnetization and thermal transport measurements at higher fields demonstrated that the AFM phase closes at the critical field $B_c\approx 10$~T, above which a phase with gapped spin excitations may appear \cite{Hong21}. 
On the other hand, Lin \textit{et al}. claimed from the magnetization and magnetic specific heat data that there is a QSL-like disordered phase in an intermediate field range $7.5\mathrm{~T} < B < 10.5\mathrm{~T}$ before entering a high-field polarized phase \cite{Lin21}. 
There is also a seemingly important anomaly in the AFM phase not identified as yet, a board hump in magnetic specific heat around 10~K present already at zero field and merging into the anomaly at $T_N$ with increasing in-plane field \cite{Yao20,Lin21}. The hump is robust against an out-of-plane field and is pronounced at higher fields, which implies a transition to the phase with a different magnetic structure. 

Although a lot of effort has been devoted to elucidate magnetic characteristics of Na$_2$Co$_2$TeO$_6$ and their relevance to Kitaev physics, fundamental aspects regarding the low-energy spin dynamics remain unresolved.  
A microscopic investigation of the phases and their evolution in magnetic fields has also been lacking. 
In this paper, we report results of comprehensive $^{23}$Na NMR measurements on polycrystalline samples of Na$_2$Co$_2$TeO$_6$, paying special attention to how magnetic phases and low-energy spin dynamics evolve with temperature and magnetic field. 
We construct the magnetic phase diagram using the microscopic quantities measured by NMR  
and demonstrate the absence of a spin-disordered phase up to 9~T.  
The nuclear spin-lattice relaxation rate $1/T_1$ is found to exhibit strong temperature and field variations especially at temperatures $T\lesssim 2T_N$. A contrasting field response of $1/T_1$ is observed above and below $T_N$. 
The measurements of $1/T_1$ in the AFM phase reveal an unconventionally large spectral weight of low-energy spin excitations which persists down to $\sim T_N/2$ and dies out below that temperature. 
As a possible origin for the qualitative change of the low-energy spin dynamics, we discuss spin-wave damping due to partially-disordered moments residing in the triple-$\mathbf{q}$ magnetic structure. 
In the paramagnetic (PM) phase, field suppression of 2D spin correlations is inferred from the field dependence of $1/T_1$. 
A dominant exchange energy is evaluated from quantitative analysis of $1/T_1$ in the high-temperature limit. 

During the course of this study, several groups have reported $^{23}$Na NMR in a single crystal of Na$_2$Co$_2$TeO$_6$~\cite{Chen21,Lee21}. Although some interesting features such as successive anomalies in the nuclear spin-spin relaxation rate $1/T_2$ and a possible signature of slow dynamics have been observed in the AFM phase, the measurements performed at a relatively low field are not intended to trace field variations of the measurable quantities. 
The present study will give additional information necessary for deeper understanding of field-dependent phenomena in this compound.

\section{Experiments}

Polycrystalline samples of Na$_2$Co$_2$TeO$_6$ were synthesized by a solid state reaction. A stoichiometric mixture of Na$_2$CO$_3$, Co$_3$O$_4$ and TeO$_2$ was pressed into pellets after fine grinding and was sintered in a preheated furnace at 860~$\mathrm{^\circ C}$ for 12~h in air. 
The samples were then cooled slowly in the furnace to room temperature over 50~h.
The final product was characterized by x-ray diffraction at room temperature and was confirmed to be a single phase with no trace of impurity. 
Temperature-dependent magnetization ($M$) measurements were carried out in a magnetic field ($B$) range of 1$-$9~T using a vibrating sample magnetometer (Quantum Design, Dynacool) under the zero field cooling (ZFC) and field cooling (FC) conditions. 

NMR measurements are performed on the $^{23}$Na nucleus (the spin $I=3/2$ and the gyromagnetic ratio $^{23}\gamma/2\pi=11.2623$~MHz/T) with a standard phase-coherent pulsed spectrometer. The $\pi/2$-$\tau$-$\pi$ two-pulse sequence was used to excite the spin-echo signals. The NMR spectra were taken by recording the spin-echo signal while sweeping the external magnetic field at a fixed frequency. $^{23}$Na nuclear spin-lattice relaxation rate $1/T_1$ was measured by an inversion recovery method at the peak position of the NMR spectrum. $1/T_1$ was determined by fitting the recovery of the spin-echo intensity $M(t)$ as a function of the time $t$ after an inversion pulse to the following stretched multi-exponential function with the exponent $\beta\;(\leq 1)$ incorporating the distribution of $T_1$ and the overlap of central ($m=1/2\leftrightarrow-1/2$) and satellite ($\pm3/2\leftrightarrow\pm1/2$) transition lines in powder samples \cite{Andrew61,Narath67,Suter98,McHenry72,Thayamballi80}: 
\begin{align}\label{eq:recovery}
	M(t) = M_\infty - M_0\sum_{k=1}^3 \alpha_k e^{-(\lambda_k t/T_1)^{\beta}}.
\end{align}
Here $M_\infty$ is the intensity at the thermal equilibrium; $M_0$ is a degree of inversion; $\{\lambda_k\}=\{1,3,6\}$ are mode eigenvalues, $\{\alpha_k\}$ are amplitudes of the corresponding modes satisfying $\sum_k \alpha_k=1$. 
Selective inversion of the populations of $m=\pm 1/2$ states (central transition) gives $\{\alpha_k\}=\{1/10,0,9/10\}$ frequently used in the literature \cite{Baek17,Zheng17,Jansa18,Takahashi19,Nagai20,Chen21,Lee21}. 
Deviation of $\beta$ from unity measures the distribution of $1/T_1$.

\section{Results}

\subsection{Magnetic susceptibility}

\begin{figure}
\includegraphics[width=7.5cm]{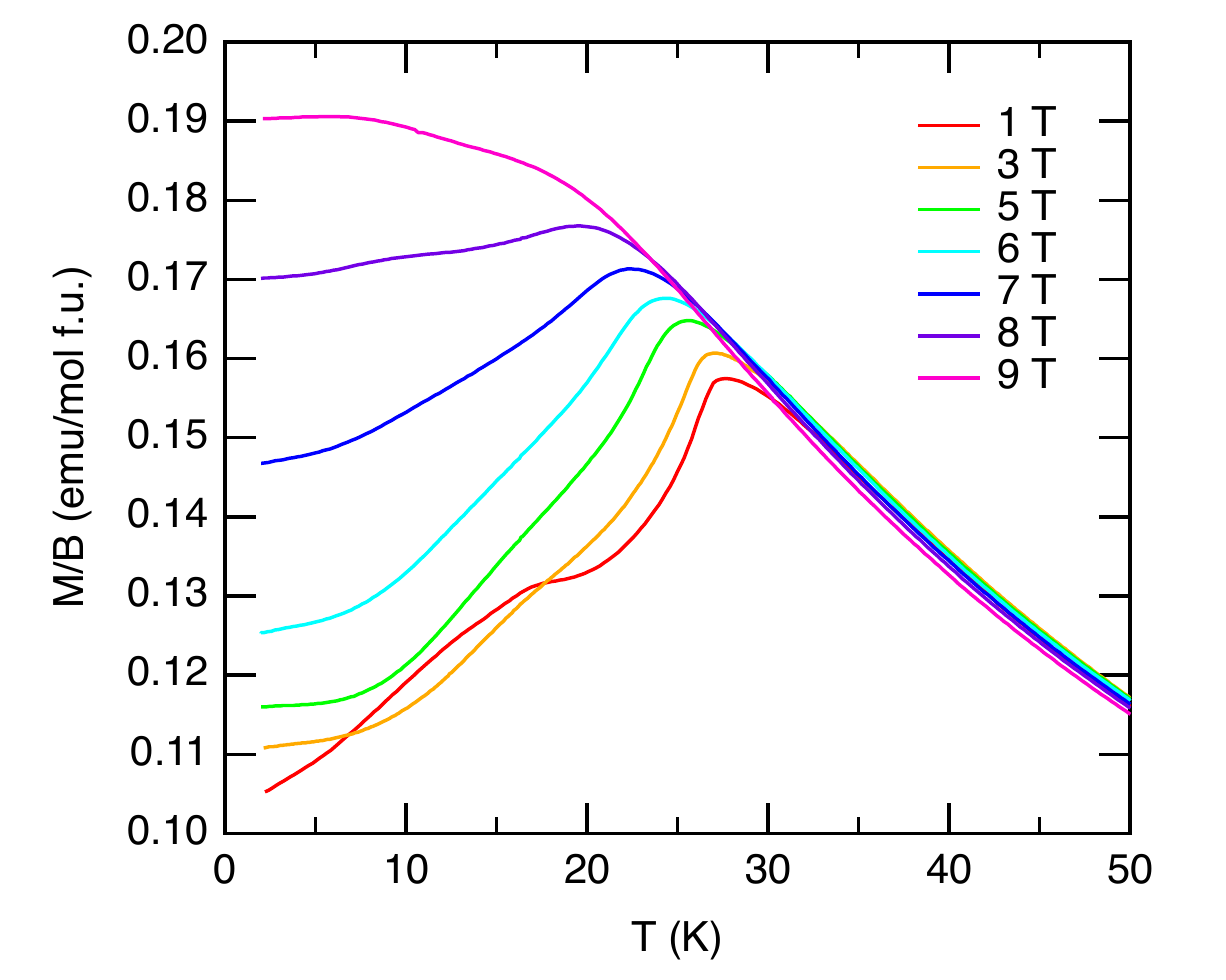}
\caption{\label{fig:sus}Temperature dependences of the magnetic susceptibility.}
\end{figure}

Figure~\ref{fig:sus} shows the temperature and field dependences of the magnetic susceptibility $M/B$ under the ZFC condition. 
No essential difference was observed between the ZFC and FC conditions except a minor difference at 1~T below $T_N$ (not shown) which may be attributed to compensation behavior of ferrimagnetism \cite{Yao20}. 
The results are in good agreement with those reported on polycrystalline samples \cite{Viciu07,Lefrancois16,Bera17} but are slightly different in the absolute magnitude from that of a single crystal because of  powder averaging of the anisotropic susceptibility \cite{Yao20,Lin21}. 
The N\'eel temperature and its field variation also agree with those in the previous reports. 
The susceptibility is essentially field independent in the PM phase, contrasting to the behavior in $\alpha$-RuCl$_3$ where it is enhanced by field at temperatures $T\lesssim 3T_N$ \cite{Yu18}. 

\subsection{\label{sec:spectrum}NMR spectrum}

\begin{figure}
\includegraphics[width=7.6cm]{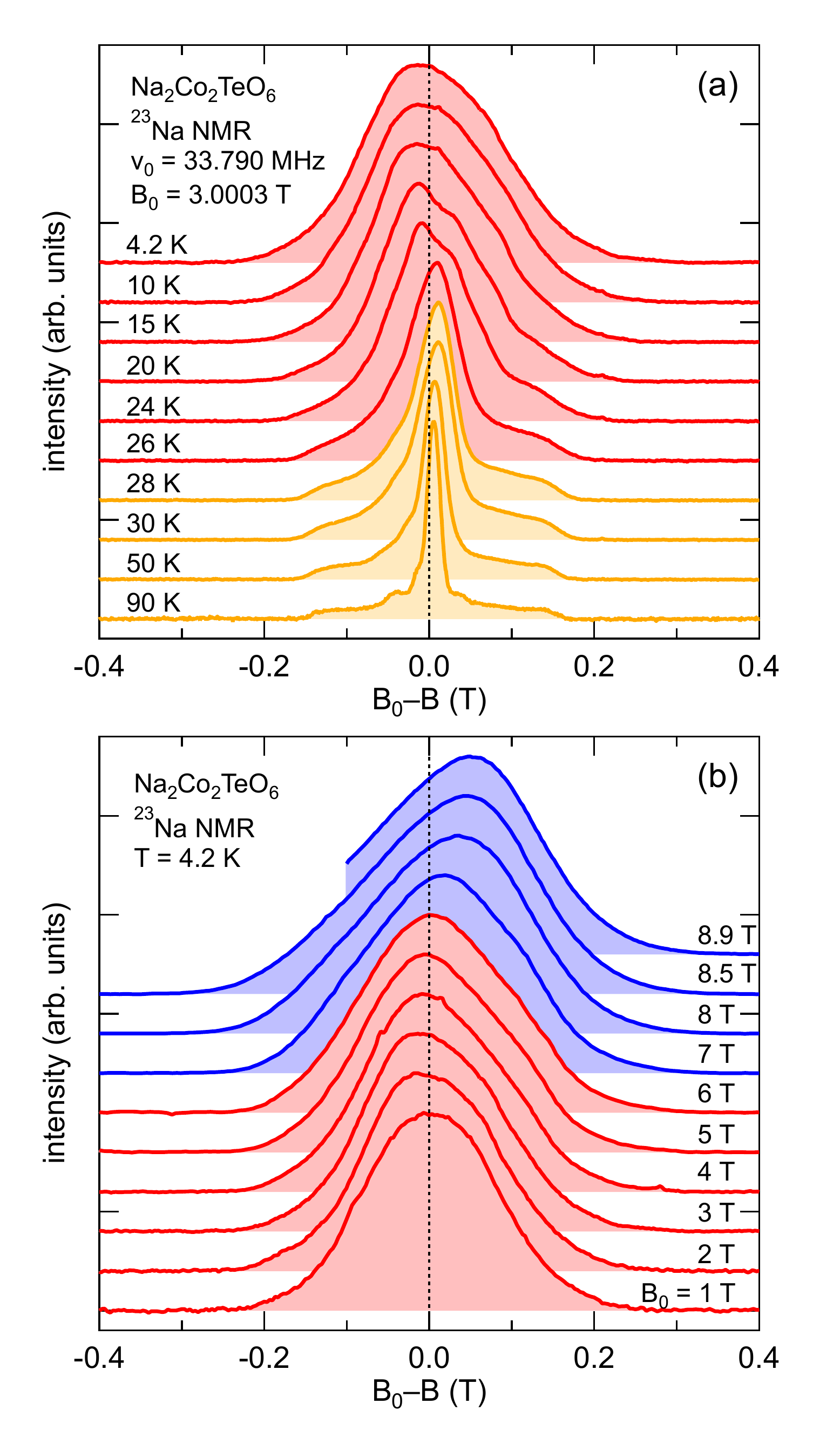}
\caption{\label{fig:spectra}(a) Temperature variation of the $^{23}$Na NMR spectrum taken at the reference field $B_0=\nu_0/\gamma\approx 3$~T. The spectra have been normalized to the maximum intensity. The spectra in the paramagnetic (antiferromagnetic) phase are colored yellow (red). (b) Field variation of the $^{23}$Na NMR spectrum at 4.2~K. The spectra with a positive peak shift ($B_0\geq 7$~T) are colored blue. The spectrum for $B_0=8.9$~T is terminated because of an upper limit of our apparatus. The dashed lines represent a reference field position $B=B_0$.}
\end{figure}

Figure~\ref{fig:spectra}(a) shows the temperature ($T$) variation of the $^{23}$Na NMR spectrum taken at a frequency $\nu_0=33.790$~MHz. The signal intensity is plotted against the field shift $B_0-B$ relative to the reference field $B_0=\nu_0/\gamma$ where $B$ is the external field. 
The spectra at high $T$ in the PM phase exhibit a quadrupolar-split powder pattern characterized by asymmetric electric field gradients (EFGs) at the Na site \cite{Abragam}. This is in clear contrast to the previous reports on single crystals, none of which observed quadrupolar satellites \cite{Chen21,Lee21}. 
The $T$-dependent magnetic broadening is also obvious (see below). From singular positions of the satellites, the principal values of the EFG tensor are determined as $^{23}\nu_\mathrm{Q} = 1.68$~MHz and $\eta = 0.49$ which are almost $T$ independent. 
We observed neither line splitting due to the presence of several crystallographic Na sites nor severe broadening or smearing of quardupolar satellites due to disorder \cite{Viciu07,Bera17,Lefrancois16}. This suggests preferential occupation of some Na site and/or a tendency of atomic ordering rather than strong disorder in Na layers. 

\begin{figure*}
\includegraphics[width=16.4cm]{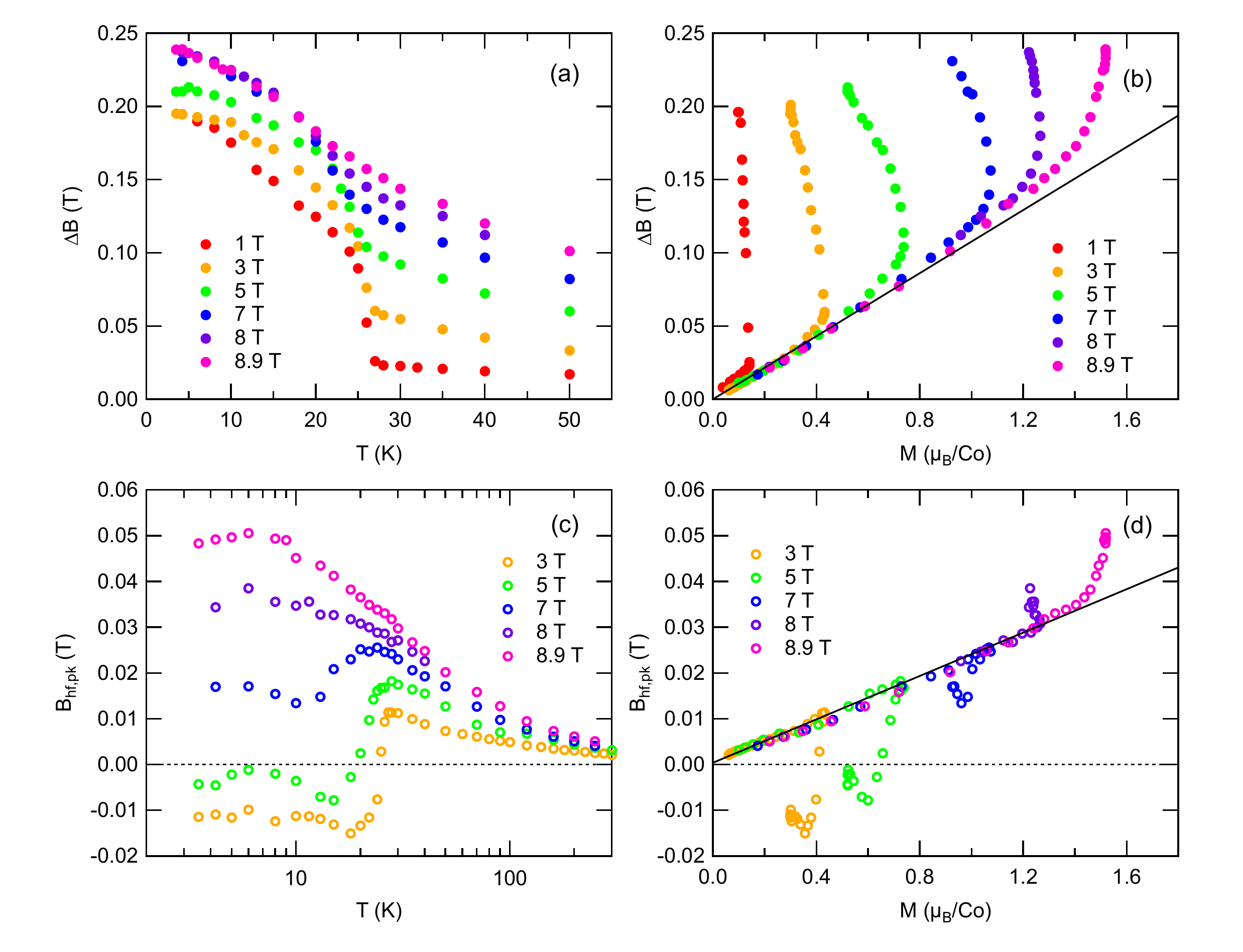}
\caption{\label{fig:fwhmBhf}(a) Temperature dependences of the fwhm $\Delta B$ of the NMR spectrum. (b) $\Delta B$ plotted against the uniform magnetization per Co atom. The solid line is a fit of the data above 50~K. The linear slope gives a coupling constant of 108 $\mathrm{mT}/\mu_B$. 
(c) Temperature dependences of the hyperfine field $B_\mathrm{hf,pk}$ at the peak position of the NMR spectrum. (d) $B_\mathrm{hf,pk}$ plotted against the uniform magnetization per Co atom. The solid line is a fit of the data above $T_N$. The linear slope gives the hyperfine coupling constant $a_\mathrm{hf}=23.7\mathrm{~mT}/\mu_B$.}
\end{figure*}

We measured the NMR spectrum at various frequencies and temperatures and determined the temperature and field dependences of the hyperfine field at the Na sites and the line width of the spectrum. 
In the following, we label the spectra and the related quantities by the reference field $B_0$. 
Figure~\ref{fig:fwhmBhf}(a) shows the $T$ dependences of the line width (fwhm) $\Delta B$ taken at various $B_0$'s. 
$\Delta B$ is also plotted in Fig.~\ref{fig:fwhmBhf}(b) as a function of uniform magnetization $M$ per Co atom with $T$ the implicit parameter. It was found that $\Delta B$ in the PM phase above 50~K is proportional to $M$ in a wide range of $B_0$. 
The proportionality constant $\Delta B/M=108\mathrm{~mT}/\mu_B$ is comparable to the root mean square of the principal values of the dipolar field tensor of $81\mathrm{~mT}/\mu_B$ calculated for the most occupied ($\sim$70\%) $12i$ site for Na. The line width in the PM phase is thus dominated by anisotropic dipolar coupling between Na and Co. 

As shown in Figs.~\ref{fig:spectra}(a) and \ref{fig:fwhmBhf}(a), sudden increase of the line width was observed on entering the AFM phase at relatively low fields $B_0\leq 5$~T. 
The low-field line widths exhibit order-parameter-like $T$ dependence, saturating at low $T$ with a slightly field-dependent value. 
The spectrum is broadened almost symmetrically about the reference field $B=B_0$, and $\Delta B$ no longer scales with $M$ [Fig.~\ref{fig:fwhmBhf}(b)]. These observations demonstrate the appearance of a staggered hyperfine field at the Na site of which magnitude is much smaller than the external field. 
The anomaly in $\Delta B$ at $T_N$ is obscured at high fields $B_0\geq 7$~T owing to magnetic broadening present already in the PM phase. The high-field line widths saturate in the low-$T$ limit to a value somewhat larger than the low-field value, possibly related to a reorientation of ordered moments with the external field. 

It should be emphasized that at low $T$ the scaling between $\Delta B$ and $M$ breaks down even at the highest field $B_0=8.9$~T. 
This is also true for the hyperfine field $B_\mathrm{hf,pk}$ at the peak position of the spectrum as will be shown below [Fig.~\ref{fig:fwhmBhf}(d)]. 
If there appeared a field-induced spin-disordered phase as suggested in Ref.~\onlinecite{Lin21}, 
both $\Delta B$ and $B_\mathrm{hf,pk}$ should scale linearly with $M$ down to the lowest temperature because of the disappearance of a staggered component of the hyperfine field. 
The upward deviation of $\Delta B$ from the $\Delta B$-$M$ scaling means that the spectrum is much broader than is expected from the uniform magnetization, demonstrating a significant contribution of the staggered hyperfine field to $\Delta B$. 
Our results thus point to the absence of a field-induced disordered phase up to $B\sim 9$~T consistent with a report that the in-plane critical field is around 10~T \cite{Hong21}.  

Figure \ref{fig:spectra}(b) shows the field evolution of the NMR spectrum at 4.2~K. The line shape as well as the line width does not change much up to $B_0=8.9$~T, which confirms the persistence of magnetic LRO. 
A shift of the peak due to slight asymmetry in the line shape is discernible at $B_0\geq 2$~T. The peak shifts to a positive side of $B_0-B$ above $B_0=7$~T, which may be associated with a magnetization jump around 6~T for the field applied perpendicular to the zigzag chains \cite{Xiao19,Yao20}. 

The magnetic shift $K$ is an important measure of the local static spin susceptibility of a magnetic ion. For nuclei with $I\geq 1$, care must be taken to correct a contribution of the quadrupole interaction to the shift of the NMR line \cite{Abragam}. We measured the field dependence of the relative line shift ${(B_0-B)/B}$ and confirmed that the second-order quadrupolar shift is negligible compared with the magnetic shift for $B_0\geq 3$~T. The field shift $B_0-B$ near the peak of the spectrum can then be taken as the hyperfine field $B_\mathrm{hf}$ at the Na sites from which the magnetic shift is defined as $K=B_\mathrm{hf}/B$.  

Shown in Fig.~\ref{fig:fwhmBhf}(c) is the $T$ dependence of $B_\mathrm{hf,pk}=B_0-B_\mathrm{pk}$ for various $B_0$'s where $B_\mathrm{pk}$ is the external field at the peak position of the spectrum. Figure~\ref{fig:fwhmBhf}(d) shows a plot of $B_\mathrm{hf,pk}$ against $M$ alternative to the conventional $K$-$\chi$ plot 
\footnote{This plot is useful in determining the hyperfine coupling constant when magnetization shows non-linear increase at high fields so that the susceptibility $\chi$ defined by $M/B$ depends on a magnetic field.}. 
The linear relation $B_\mathrm{hf,pk}=a_\mathrm{hf}M$ holds in the PM phase in a wide range of $B_0$. 
The slope $a_\mathrm{hf}=23.7\mathrm{~mT}/\mu_B$ is by definition the hyperfine coupling constant between Na and Co. The small and positive value of $a_\mathrm{hf}$ suggests that the Na-$3s$ orbital is polarized a little due to spin transfer from the neighboring Co atoms. 
Breakdown of the scaling between $B_\mathrm{hf,pk}$ and $M$ in the AFM phase is signaled by deflection of the plot from the straight line. The deviation of the low-$T$ data at the highest field $B_0=8.9$~T from the scaling confirms the absence of a spin-disordered phase up to $B\sim 9$~T. 

Another interesting behavior in the AFM phase is a field-dependent shift of $B_\mathrm{hf,pk}$ accompanying the sign change [Fig.~\ref{fig:fwhmBhf}(c)]. 
As noted above, this is due to asymmetry in the line shape that reflects the distribution of staggered hyperfine fields. The negative shift of the peak at low fields may be associated with domains of net magnetic moment directed opposite to the external field which exist in ferrimagnets showing compensation behavior. 
The positive shift at high fields is probably caused by a decrease in the number of such domains as well as an increasing contribution of uniform magnetization to the hyperfine field. 

\begin{figure}
\includegraphics[width=7.8cm]{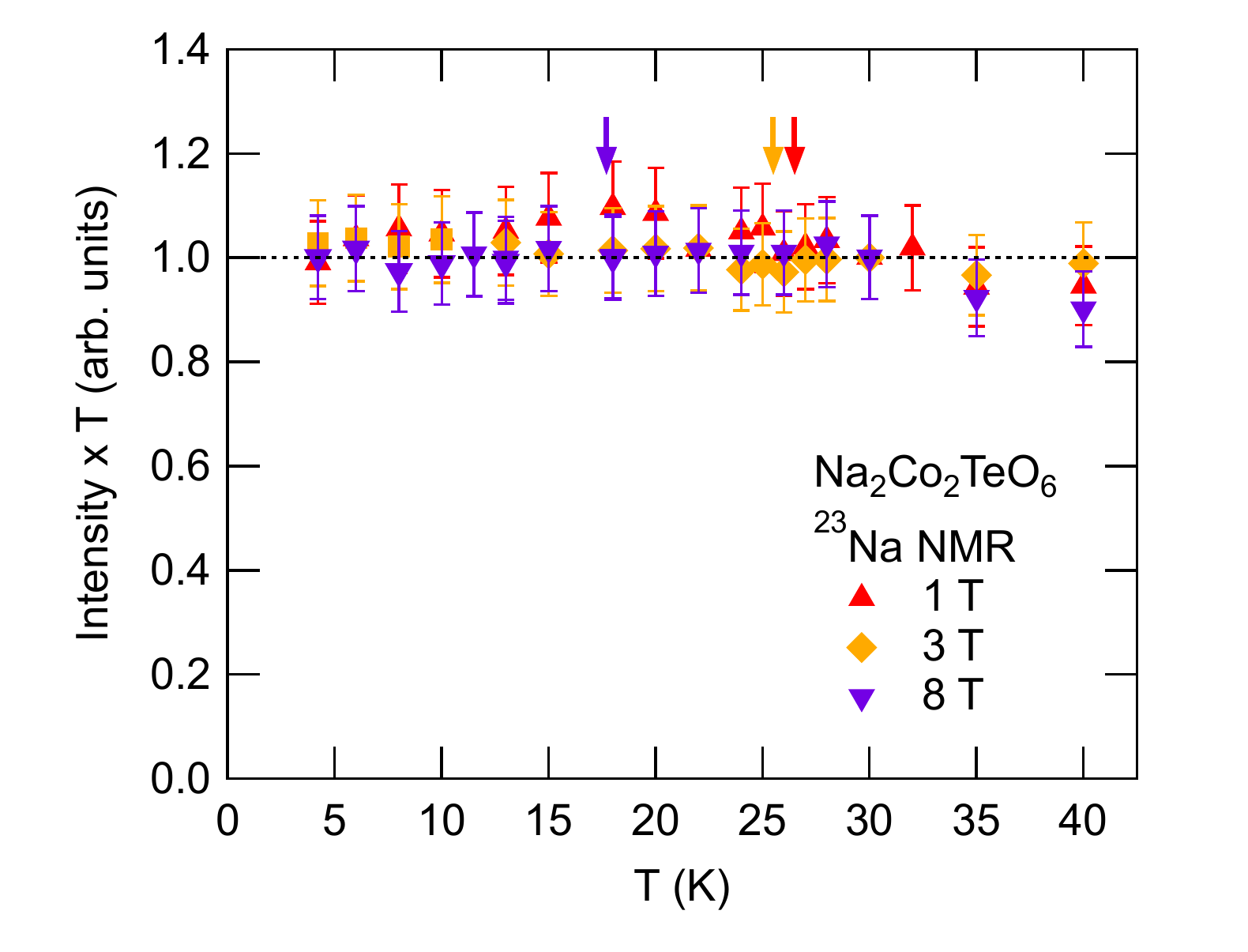}
\caption{\label{fig:intensity}Temperature dependences of the integrated intensity of the NMR spectrum multiplied by temperature. The intensities have been normalized at 30~K. The arrows indicate $T_N$ at the reference field of the corresponding color.}
\end{figure}

Integrated intensity of the NMR spectrum is one of the essential quantities to be measured in frustrated magnets because it is highly sensitive to slow dynamics of which existence is manifested by the loss of signal intensity, or wipeout \cite{MacLaughlin76,Mendels00,Olariu06}. 
Recently, the slow dynamics of Co spins in the AFM phase has been suggested from partial wipeout of the $^{23}$Na NMR signal below $T_N$ \cite{Chen21}. 
To avoid an artifact of finite separation $\tau$ between rf exciting pulses in determining the integrated intensity, we measured spin-echo decay at the peak position of the spectrum and corrected the intensity by extrapolating it to $\tau=0$. 
The results are shown in Fig.~\ref{fig:intensity} where we plotted the corrected intensity multiplied by $T$, a factor arising from nuclear paramagnetism, as a function of $T$.  
Unlike the previous report, we observed no anomaly around and below $T_N$; 
the integrated intensity is essentially $T$ independent down to 4.2~K. This clearly shows the absence of slow dynamics in the measured $T$ range 
\footnote{We detected a dip in the uncorrected intensity (typically taken at $\tau=40\mathrm{~\mu s}$) around $T_N$ as reported in Ref.~\onlinecite{Chen21} which is recovered by the extrapolation process}. 
The apparent decrease in intensity around $T_N$ reported in Ref.~\onlinecite{Chen21} could be caused by an insufficient correction of shortening of spin-spin relaxation time $T_2$ accompanied by critical slowing down. 
The unrecovered signal loss below $T_N$ might also be an artifact of insufficient integration range resulting from the large spectral broadening due to magnetic LRO.

\subsection{\label{sec:T1}Nuclear spin-lattice relaxation rate}

\begin{figure*}
\includegraphics[width=16.6cm]{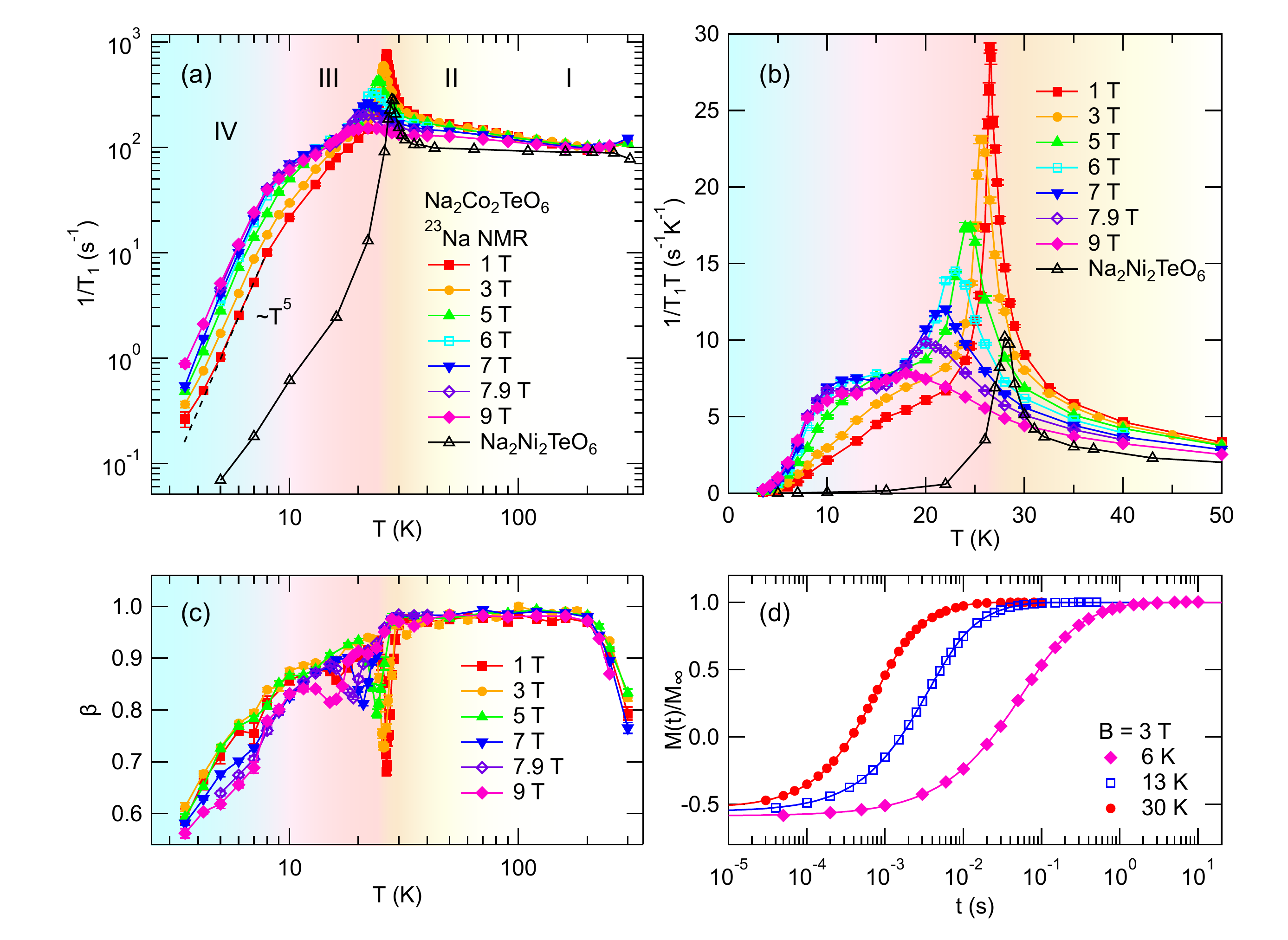}
\caption{\label{fig:T1}(a) Temperature dependences of $1/T_1$. The data of Na$_2$Ni$_2$TeO$_6$ were taken from Ref.~\onlinecite{yitoh15}. Four temperature regions are labeled with numbers from I to IV. 
The dashed line shows a power law $1/T_1\propto T^5$. 
(b) Temperature dependences of $1/T_1T$ at low temperatures. 
(c) Temperature dependences of the stretching exponent $\beta$ of the magnetization recovery. (d) Examples of the magnetization recovery taken at $B=3$~T. The solid lines are fits to the stretched multi-exponential function Eq.~\eqref{eq:recovery}.}
\end{figure*}

Figure~\ref{fig:T1}(d) shows examples of the recovery of $^{23}$Na spin-echo intensity $M(t)$. The data are well fitted by Eq.~\eqref{eq:recovery}, which demonstrates reliability of the analyses. 
The $T$ and $B$ dependences of $1/T_1$ at the Na sites are shown in Fig.~\ref{fig:T1}(a), and the low-$T$ close-up in the form of $1/T_1T$ in Fig.~\ref{fig:T1}(b). The $^{23}$Na $1/T_1$ in Na$_2$Ni$_2$TeO$_6$ with a similar N\'eel temperature $T_N\approx 26$~K but with a slightly different stacking sequence of honeycomb layers are shown for comparison \cite{yitoh15}. The $^{23}$Na $1/T_1$ in Na$_2$Co$_2$TeO$_6$ is characterized by a strong variation not only with temperature but with magnetic field. 
This makes a striking contrast to the macroscopic susceptibility which depends hardly on magnetic field in the PM phase up to 9~T [Fig.~\ref{fig:sus}]. 
The four characteristic temperature regions are identified: \textit{Region I}, above 50$-$60~K ($\sim 2T_N$) where $1/T_1$ is nearly $B$ independent and the $T$ dependence is relatively weak; \textit{Region II}, $T_N<T\lesssim 2T_N$ where $1/T_1$ increases toward $T_N$ and the $B$ dependence becomes noticeable; \textit{Region III}, $T_N/2\lesssim T<T_N$ where $1/T_1$ decreases with decreasing $T$ but remains enhanced, still exhibiting strong field evolution; and \textit{Region IV}, $T\lesssim T_N/2$ where a rapid decrease of $1/T_1$ is observed regardless of the field.  
On the $B$ dependence, $1/T_1$ shows a contrasting response to magnetic fields above and below $T_N$; $1/T_1$ is suppressed by field above $T_N$ whereas it is enhanced below $T_N$. 

Let us inspect first the $T$ dependence of $1/T_1$ at the lowest field. 
$1/T_1$ measured at $B=1$~T depends only weakly on $T$ above about 60~K, approaching an almost $T$- and $B$-independent value of $1/T_{1\infty}\sim 100\mathrm{~s^{-1}}$ as usually observed in magnets with exchange-coupled local moments. 
On decreasing $T$ across 60~K, $1/T_1$ starts to increase due to the development of short-range spin correlations consistent with the neutron diffuse scattering \cite{Bera17}. 
This is followed by a divergent increase of $1/T_1$ on approaching $T_N$ below about 30~K, an indication of three-dimensional (3D) critical slowing down.
$1/T_1$ takes a maximum at $T_N\approx 26.5$~K and then decreases rapidly on cooling. However, the decrease of $1/T_1$ is rather gradual compared with that in a conventional antiferromagnet in which $1/T_1$ often decreases many orders of magnitude not far below $T_N$ as observed in Na$_2$Ni$_2$TeO$_6$. $1/T_1$ remains in the same order as in the PM phase, which indicates residual low-energy spin excitations in Region III. 
This is closely related to the result of single-crystal INS that the low-energy spectral weight survives down to $\sim$14~K without forming a spin-wave mode and a gap in the excitation spectrum \cite{Chen21}. 
The absence of an upturn of thermal conductivity on entering the AFM phase \cite{Hong21} may have a close connection to the residual low-energy spin excitations. 
On further cooling below 13$-$15~K to Region IV, $1/T_1$ decreases rapidly over two decades. 
The $T$ dependence of $1/T_1$ is approximated by a power law $1/T_1\propto T^n$ with $n\approx 5$ predicted for the three-magnon process at $T\gg\Delta$ rather than an activation law expected at $T\ll\Delta$ where $\Delta$ is the gap in the spin-wave spectrum \cite{BP}. 

Our result at $B=1$~T is in good agreement with the previous report of $1/T_1$ for a single crystal by Chen \textit{et al}. taken at $B=0.75$~T with $\mathbf{B}\parallel a^*$ from 15 to 60~K \cite{Chen21}.
Minor differences in the absolute magnitude of $1/T_1$ may be attributed to our use of the stretched exponential function Eq.~\eqref{eq:recovery} as well as of polycrystals 
\footnote{The stretched exponential function used in Ref.~\onlinecite{Chen21} is slightly different from conventional ones: the exponential terms are like $\exp[-6(t/T_1)^\beta]$ rather than $\exp[-(6t/T_1)^\beta]$. If this form of stretched exponentials is applied to our recovery, the resulting $1/T_1$ agrees quantitatively with their $1/T_1$.}. 
A more extensive single-crystal study has been reported by Lee \textit{et al}. who measured both $1/T_1$ and $1/T_2$ at $B\sim 3.1$~T with $\mathbf{B}\parallel c$ and $\perp c$ up to room temperature \cite{Lee21}. 
The $T$ dependence of their $1/T_1$ above $\sim$10~K is qualitatively similar to ours at 3~T, but the absolute values are larger by a factor of $\sim$2. 
The discrepancy becomes progressively greater below $\sim$10~K in Region IV, leading to a moderate $T$ variation of their $1/T_1$ described by a power law with a smaller exponent $n\sim 3$. 

In the AFM phase where the powder NMR spectrum is largely broadened due to the staggered hyperfine field, 
we measured $1/T_1$ at several positions of the spectrum other than the peak position 
and found that the $1/T_1$ differs by at most 10\% in parallel with the result of $1/T_1$ for the split peaks in Ref.~\onlinecite{Lee21}. 
We also performed the inverse Laplace transform analysis of magnetization recovery \cite{Singer20,Arsenault20} in order to see whether the stretched exponential analysis captures the true distribution of $1/T_1$. 
The distribution function of $1/T_1$ similar to that expected for the stretched exponential recovery \cite{Johnston05,Johnston06} was obtained as detailed in the Appendix. 
This justifies our phenomenological analysis of $1/T_1$ using Eq.~\eqref{eq:recovery} and rules out a possibility of the inequivalent Na sites showing a distinct $T$ variation of $1/T_1$ from the one displayed in Fig.~\ref{fig:T1}.  

The origin of the large discrepancy between our $1/T_1$ and that reported in Ref.~\onlinecite{Lee21} especially in the behavior in Region IV is unclear. 
This is partly because the authors of Ref.~\onlinecite{Lee21} did not give fundamental information on the recovery such as the $T$ dependence of $\beta$ to be compared with ours and the recovery itself from which $1/T_1$ is extracted. 
A possible origin of the discrepancy is a different degree of atomic disorder in Na layers between the samples; their single-crystal sample may have stronger disorder than our polycrystals because quadrupolar satellites are missing in their NMR spectrum 
\footnote{Vanishingly small quadrupolar splitting or negligible quadrupole coupling seems improbable for their missing of quadrupolar satellites because they can fit the magnetization recovery using the form for a quadrupolar-split center line.}. 
Disorder in Na layers might change low-energy spin excitations by modulating interlayer coupling and/or by introducing bond randomness in Co layers, affecting $1/T_1$ effectively at low $T$ where the intrinsic $1/T_1$ falls off due to magnetic LRO. 
We should also point out that in the PM phase the Redfield contribution ($T_1$ process) to $1/T_2$ for $\mathbf{B}\parallel c$ evaluated from their $1/T_1$ for $\mathbf{B}\parallel c$ and $\perp c$ exceeds the observed $1/T_2$ for $\mathbf{B}\parallel c$, which is impossible for the quadrupolar-split center line \cite{Walstedt67,Auler96}. 
To resolve the conflicts, single-crystal NMR measurements are under way and will be reported in a future publication. 

Let us go back to our results at higher fields. 
Increasing magnetic field suppresses $1/T_1$ above $T_N$, most drastically around $T_N$ where the peak gets broadened and is shifted to lower $T$. This suggests field-induced suppression of short-range spin correlations.  
In contrast, $1/T_1$ is enhanced by field below $T_N$ to exhibit a shoulder-like anomaly around 10~K.
The anomaly appears as a broad hump in a plot of $1/T_1T$ and is most pronounced at 6$-$7 T  [Fig.~\ref{fig:T1}(b)], which indicates field enhancement of the spectral weight remaining at low energies. 
The field enhancement of $1/T_1$ cannot be ascribed to defect spin fluctuations because in that case $1/T_1$ would be suppressed by magnetic fields \cite{McHenry72,Kitagawa18}. 
On the other hand, the magnetic field does not affect much the $T$ dependence of $1/T_1$ in Region IV; $1/T_1$ shows an approximate $T^5$ law even at 9~T. 
There is no indication of strong field enhancement of the low-energy excitations. This corroborates the absence of a spin-disordered phase up to 9~T.

The rounding of the peak of $1/T_1$ around $T_N$ and the appearance of a broad hump in $1/T_1T$ around 10~K in magnetic fields resemble the behavior of magnetic specific heat $C_m/T$ \cite{Yao20,Lin21}. 
This suggests that $1/T_1$ probes excitations governing the magnetic specific heat.  
The resemblance between $1/T_1T$ and $C_m/T$ also suggests that the rounded peak of $1/T_1$ in magnetic fields is not an artifact due to the use of powder samples but an intrinsic property of this compound 
\footnote{Since the magnetic response of Na$_2$Co$_2$TeO$_6$ is insensitive to the field applied perpendicular to the honeycomb planes, it seems reasonable to consider that the field dependence measured in powders is governed by in-plane field responses. This is further supported by the fact that in powders a grain with the field lying in the honeycomb plane is found more frequently than a grain with the field normal to the plane, because the probability of a field making an angle $\theta$ with respect to the direction normal to the plane is proportional to $\sin\theta$.}. 

The stretching exponent $\beta$ in Eq.~\eqref{eq:recovery} also includes valuable information on the spin dynamics. 
As shown in Fig.~\ref{fig:T1}(c), $\beta$ is nearly $T$- and $B$-independent from 30 to 200~K and takes a value close to unity, which indicates a nearly uniform relaxation process. 
A rapid decrease of $\beta$ above 200~K implies the appearance of additional relaxation channels possibly related to the spin-orbit excited state lying 22$-$23~meV above the ground state \cite{Songvilay20,KimC22}. 
This provides an indirect support for the spin-orbital entangled state of Co$^{2+}$ in Na$_2$Co$_2$TeO$_6$. 
$\beta$ decreases steeply below 30~K and takes a local minimum around $T_N$. 
A large distribution of $1/T_1$ close to $T_N$ may result from strong temperature and field-orientation dependence of $1/T_1$. 
While modest inhomogeneity of $1/T_1$ ($\beta\gtrsim 0.9$) is found above $\sim$10~K, 
inhomogeneous relaxation prevails very rapidly below $\sim$10~K. 
This suggests a qualitative change of the spin dynamics on entering Region IV.

\subsection{Phase diagram}

\begin{figure}
\includegraphics[width=8.4cm]{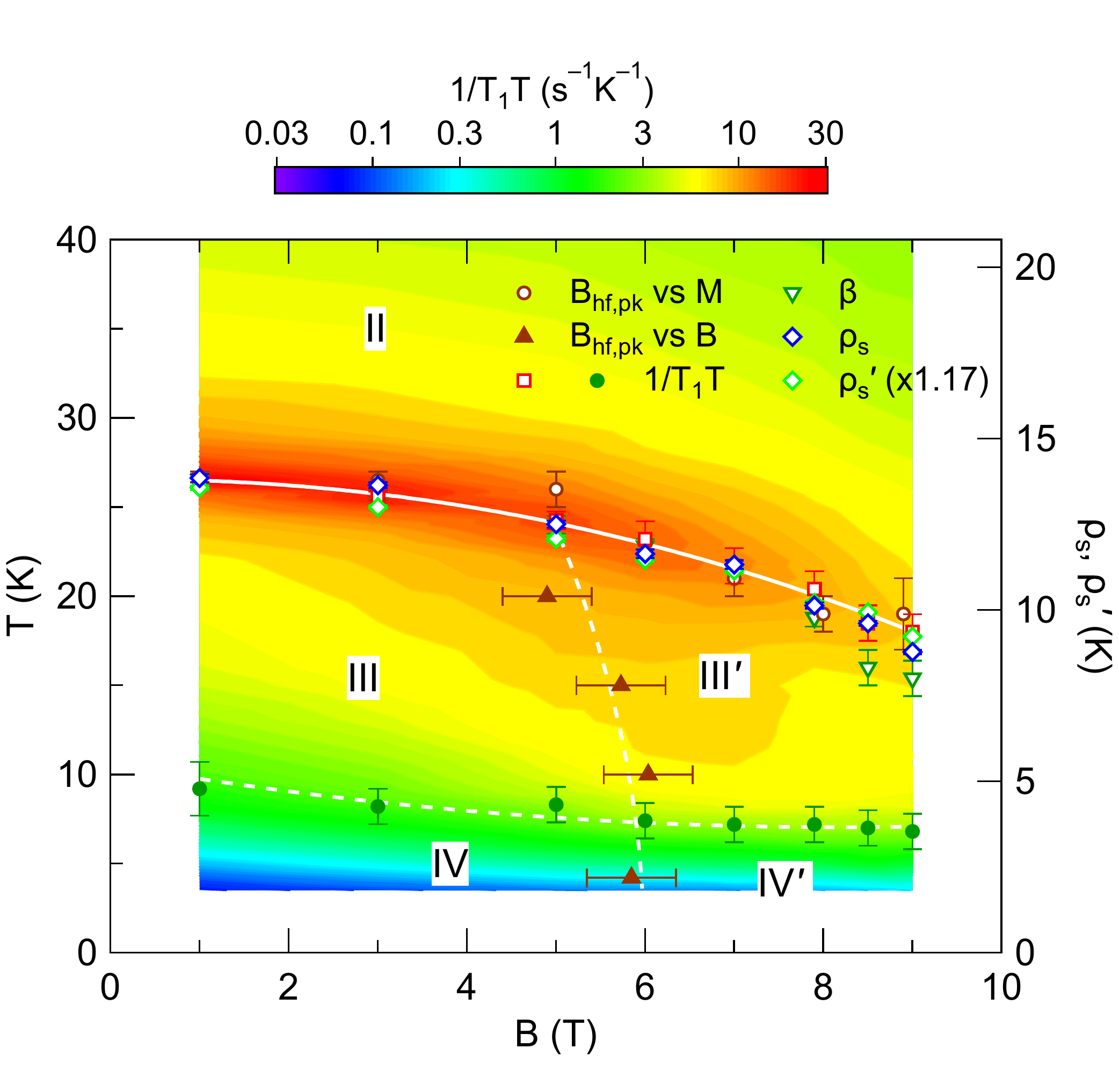}
\caption{\label{fig:phase}Magnetic phase diagram constructed from the quantities measured by $^{23}$Na NMR. A contour plot of $1/T_1T$ is also shown for comparison. The primed letters represent the high-field ordered phase with a possible reversal of the in-plane canted moment \cite{Yao20}. The lines between the phases and the regions are guides to the eyes. Right axis: field dependence of the spin stiffness constants $\rho_s$ and $\rho_s^{\,\prime}$ determined from the renormalized-classical analyses of $1/T_1$ (see text). $\rho_s^{\,\prime}$ is scaled up by multiplying a factor of 1.17 to demonstrate that the field dependence of $\rho_s$ and $\rho_s^{\,\prime}$ is identical. The scale of the right axis is adjusted for $\rho_s$ at 1~T to coincide with the point for $1/T_1T$ at 1~T plotted with respect to the left axis.}
\end{figure}

The magnetic phase diagram is constructed from the $T$- and $B$-dependences of various quantities measured by NMR. The result is summarized in Fig.~\ref{fig:phase} together with a contour plot of $1/T_1T$ for comparison. $T_N$ may be determined in several ways: the temperature below which the $B_\mathrm{hf,pk}$-$M$ scaling breaks down [Fig.~\ref{fig:fwhmBhf}(d)], the temperature at which $1/T_1T$ takes a maximum, and the temperature at which $\beta$ takes a local minimum. 
All of these agree within experimental accuracies. 
Determination of the boundary between Regions III and IV seems more difficult because of a gradual nature of the transition. We tentatively adopt the temperature at which the derivative of $1/T_1T$ takes a local maximum. 
The field above which $B_\mathrm{hf,pk}$ shifts to a positive side [Fig.~\ref{fig:spectra}(b)] constitutes another boundary dividing low- and high-field AFM phases, 
although the small shift of $B_\mathrm{hf,pk}$ relative to the broad powder spectrum leads to large errors in the boundary field. 
The phase diagram is in good agreement with that determined from the macroscopic quantities \cite{Yao20,Hong21,Lin21}. 

Our main finding is the existence of two distinct temperature regions in the AFM phase where Co spins exhibit contrasting low-energy dynamics. In Region III, the spin excitation spectrum has a significant low-energy weight that is enhanced strongly with magnetic field. 
In contrast, Region IV is likely described by spin-wave excitations. 
As mentioned above the transition between the two regions is gradual and is possibly a crossover rather than a phase transition with critical dynamics. 

A transition to the high-field ordered phase (Regions III$^\prime$ and IV$^\prime$) 
was detected around 6~T by the static quantity $B_\mathrm{hf,pk}$, reflecting the magnetization jump associated with a reversal of canting moments \cite{Yao20}. 
The low-energy spin dynamics is essentially unchanged by this transition, although the field response of $1/T_1$ is somewhat weakened. A spin-disordered phase does not appear up to 9~T.

\section{\label{sec:anaT1}Analysis of spin-lattice relaxation rate}

\subsection{High-temperature limit}

The exchange interactions between Co spins have been evaluated by several groups via powder INS techniques, but the results are not settled yet \cite{Songvilay20,Lin21,KimC22,Samarakoon21}. Here we present an independent evaluation of the dominant interaction strength from the high-$T$ limiting value of $^{23}$Na $1/T_1$ which is helpful in justifying a proper energy scale of this compound. 

In the PM phase at temperatures much higher than the exchange interactions, the spin dynamics is modeled by Gaussian random modulation of individual spins under the influence of exchange-coupled neighbors. The characteristic (exchange) frequency $\omega_e$ is determined solely by the interaction strengths, leading to $T$-independent $1/T_1$ \cite{Moriya-I,Moriya-II}. 
In the presence of both dipolar and isotropic transferred hyperfine coupling, $1/T_1$ in the high-$T$ limit is given as a sum of two contributions; 
\begin{align}
	\frac{1}{T_{1\infty}} &= \frac{1}{T_{1\infty,\mathrm{dip}}} + \frac{1}{T_{1\infty,\mathrm{tr}}}.
\end{align}
The dipolar contribution is expressed as 
\begin{align}\label{eq:T1dip}
	\frac{1}{T_{1\infty,\mathrm{dip}}} &= \sqrt{2\pi} (\gamma g\mu_B)^2 
	\sum_l r_l^{-6}\frac{2S(S+1)}{3\omega_e},
\end{align}
where $r_l$ is a distance between the nucleus and the $l$-th electron spin, $g$ is the $g$-factor. $\omega_e$ is given in terms of the exchange energy $J_{ij}$ between $i$-th and $j$-th electron spins and the number of $j$-th spins $z_j$ coupled to the $i$-th spin as 
\begin{align}\label{eq:wex}
	\omega_e^2 &= \frac{2}{3}S(S+1)\sum_j z_j\biggl(\frac{J_{ij}}{\hbar}\biggr)^2. 
\end{align}
For the contribution of the transferred hyperfine coupling which we assume to come from the nearest-neighbor electron spins, we have 
\begin{align}\label{eq:T1tr}
	\frac{1}{T_{1\infty,\mathrm{tr}}} &= \sqrt{\frac{\pi}{2}} (\gamma g\mu_Ba_\mathrm{hf})^2 
	\frac{2S(S+1)}{3z_\mathrm{n}\omega_e}.
\end{align}
Here $a_\mathrm{hf}$ is the hyperfine coupling constant, $z_\mathrm{n}$ is the number of nearest-neighbor spins coupled to the nucleus. We take the structural model of Xiao \textit{et al}. \cite{Xiao19} for simplicity and put $z_\mathrm{n}=4$. 

Evaluating the sum $\sum_l r_l^{-6}$ in Eq.~\eqref{eq:T1dip} within a sphere of radius $100\mathrm{~\AA}$ and using the value of $a_\mathrm{hf}$ determined from the $B_\mathrm{hf,pk}$-$M$ scaling in Fig.~\ref{fig:fwhmBhf}(d), we obtain the ratio of the two contributions as $T_{1\infty,\mathrm{dip}}/T_{1\infty,\mathrm{tr}} = a_\mathrm{hf}^2/2 z_\mathrm{n}\sum_l r_l^{-6} \approx 0.02$. 
Hence we neglect the contribution $1/T_{1\infty,\mathrm{tr}}$ and go on to evaluate $\omega_e$ using Eq.~\eqref{eq:T1dip}. 
Adopting the isotropic value $g=4.33$ for the $S=j_\mathrm{eff}=1/2$ manifold of Co$^{2+}$ for simplicity \cite{AB}, we finally obtain $\omega_e = 4.1\times 10^{12}\mathrm{~s^{-1}}$ from the observed value $1/T_{1\infty}=100\mathrm{~s^{-1}}$. 

Since the contributions of $J_{ij}$'s to $\omega_e$ are all additive, the maximum value of some $J_{ij}$ may be estimated by neglecting all the other contributions. By putting $J_{ij}=0$ other than the nearest-neighbor Heisenberg coupling $J$ and by putting $z_1=3$, we get $\abs{J} = 26$~K (2.2~meV). Note that the sign of $J_{ij}$ cannot be determined by the present analysis. 
In estimating the Kitaev coupling $K$, putting $z_1=1$ gives $\abs{K} = 44$~K (3.8~meV). 
The obtained values of $\abs{J}$ and $\abs{K}$ fall in the same order as those evaluated from the INS experiments \cite{Songvilay20,KimC22,Samarakoon21,Lin21,Sanders22}.
According to the microscopic model of the exchange interactions in Na$_2$Co$_2$TeO$_6$ \cite{Liu21}, this value of $K$ sets the energy scale $t^2/U\approx 1.1$~meV where $t$ is the hopping amplitude and $U$ is Coulomb repulsion defined in Refs.~\onlinecite{Liu20,Liu21}.

\subsection{Region with short-range spin correlations}

At temperatures $T_N<T\lesssim 2T_N$ (Region II), the neutron diffuse scattering experiment has revealed that short-range spin correlations develop within the honeycomb planes \cite{Bera17}. The field suppression of $^{23}$Na $1/T_1$ thus suggests a reduction of the in-plane spin correlation length in magnetic fields. In order to quantify the $B$ dependence of $1/T_1$ from such a standpoint, we analyze the $T$ dependence of $1/T_1$ based on a description of 2D Heisenberg antiferromagnets (HAFs) in terms of the quantum non-linear $\sigma$ model.  
Similar analysis has recently been applied to $^{23}$Na $1/T_1$ in a single crystal of Na$_2$Co$_2$TeO$_6$ at a moderate field \cite{Lee21}.  

We assume that the system is in the renormalized-classical (RC) regime of 2D HAFs.
For the system with collinear order \cite{CHN,Chakravarty90,Imai93}, the correlation length $\xi$ grows exponentially with decreasing $T$ in the RC regime as $\xi\propto \exp(2\pi\rho_s/T)$ where $\rho_s$ is the spin stiffness constant 
\footnote{Including the correction term of $O(T/2\pi\rho_s)$ for $\xi$ introduces a minor change of parameters ($\sim$10\% reduction of $\rho_s$ for example) but does not change the results qualitatively}. 
The spin-lattice relaxation rate is given as $1/T_1 \propto T^{3/2}\xi$, so that
\begin{gather}\label{eq:RCC}
	\frac{1}{T_1} \propto T^{3/2}\exp(2\pi\rho_s/T).
\end{gather}
Considering the possibility of triple-$\mathbf{q}$ order \cite{Chen21}, we also examine the non-linear $\sigma$ model developed for the system with noncollinear order \cite{Azaria92,Chubukov94a,Chubukov94b}. For this type of order, $\xi\propto T^{-1/2} \exp(4\pi\rho_s^{\,\prime}/T)$ and $1/T_1\propto T^{7/2}\xi$, giving \cite{yitoh09}
\begin{gather}\label{eq:RCNC}
	\frac{1}{T_1} \propto T^3\exp(4\pi\rho_s^{\,\prime}/T). 
\end{gather}
Here $\rho_s^{\,\prime}$ is the corresponding spin stiffness constant. 
\begin{figure}
\includegraphics[width=8.2cm]{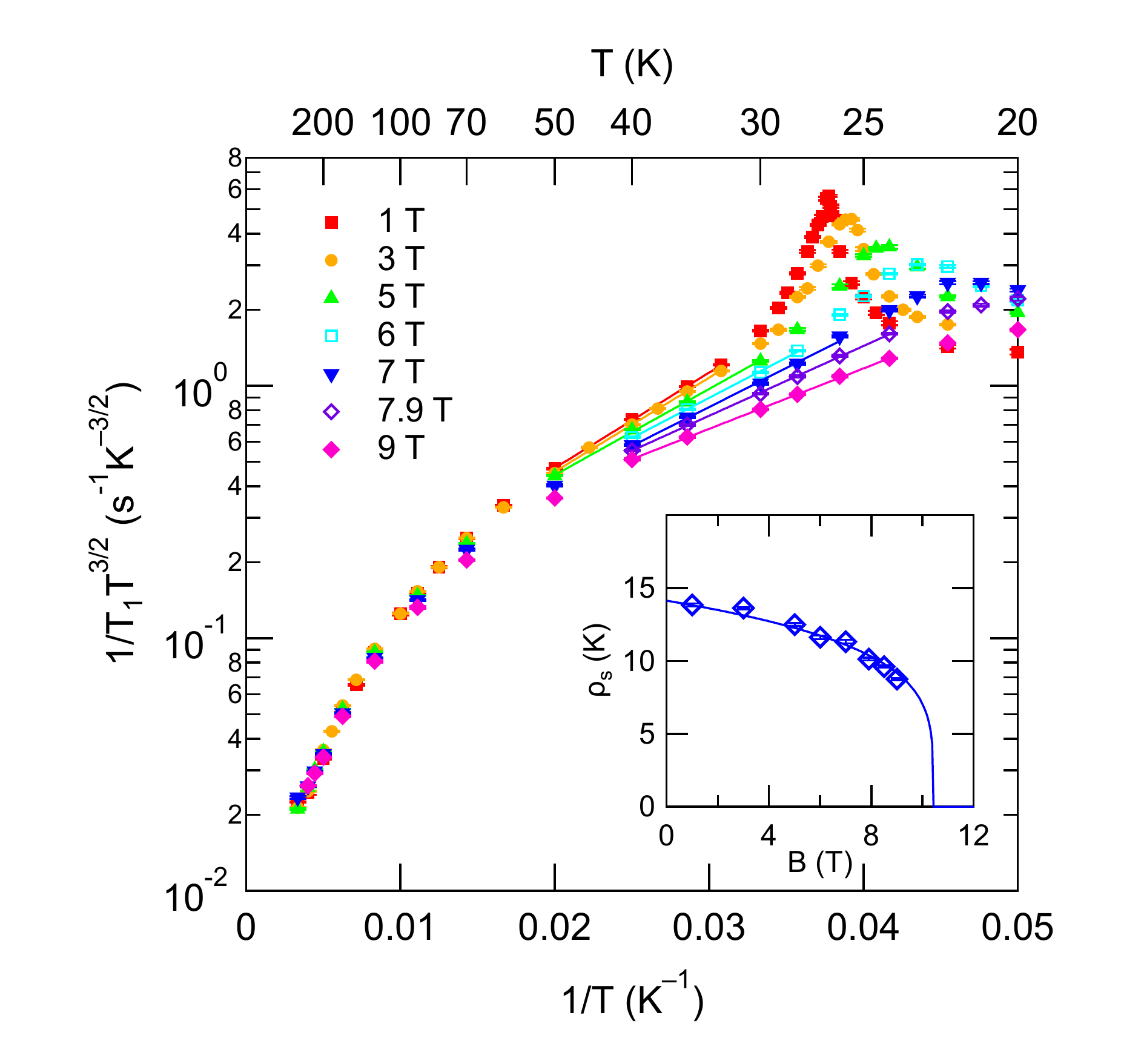}
\caption{\label{fig:RC}Semi-logarithmic plots of $1/T_1T^{3/2}$ versus $1/T$. The lines are fits to the renormalized-classical form $1/T_1 \propto T^{3/2}\exp(2\pi\rho_s/T)$. Field dependence of $\rho_s$ is shown in the inset together with the fitting to the power law $\rho_s\propto (B_c-B)^p$.}
\end{figure}

Figure \ref{fig:RC} shows $1/T_1T^{3/2}$ plotted against the inverse temperature $1/T$ in a semi-logarithmic scale. 
$1/T_1$ is found to obey the scaling relation Eq.~\eqref{eq:RCC} below 50 to 32~K (40 to 24~K at high fields) which ensures that the system is indeed in the RC regime. The scaling form Eq.~\eqref{eq:RCNC} for noncollinear order is also satisfied in the same $T$ range (not shown). 
$\rho_s$ decreases monotonically with increasing field as shown in the inset of Fig.~\ref{fig:RC}, showing a tendency to vanish at 10$-$12~T. 
The $B$ dependence of $\rho_s^{\,\prime}$ is identical to that of $\rho_s$ and is scaled with a multiplicative factor $\rho_s/\rho_s^{\,\prime}=1.17$ [Fig.~\ref{fig:phase}]. 

The spin stiffness constant characterizes the rigidity of an ordered state and is nonzero only in a phase with magnetic LRO. In the quantum non-linear $\sigma$ model for 2D HAFs, it is renormalized by quantum fluctuations and vanishes on approaching a quantum critical point, beyond which a quantum disordered state appears \cite{CHN}. The field-induced reduction of $\rho_s$ and $\rho_s^{\,\prime}$ thus suggests the system getting closer to the quantum disordered phase present above a certain critical field. 
This also implies that the high-field phase of Na$_2$Co$_2$TeO$_6$ is not a QSL but a partially polarized phase showing bosonic excitations.  
As a rough estimate of the critical field $B_c$, we fitted the $B$ dependence of $\rho_s$ and $\rho_s^{\,\prime}$ to the power law $\rho_s,\rho_s^{\,\prime}\propto (B_c-B)^p$, getting $B_c=10.4$~T and $p=0.22$. The obtained value of $B_c$ agrees well with the value $B_c\approx 10$~T determined from the magnetization measurement \cite{Hong21}, although the agreement seems fortuitous considering the lack of our data closer to $B_c$.  

The $B$ dependence of $\rho_s$ and $\rho_s^{\,\prime}$ is also shown in Fig.~\ref{fig:phase} for comparison with the phase diagram. 
Surprisingly enough, $\rho_s$ and $\rho_s^{\,\prime}$ trace $T_N$ as a function of magnetic field with appropriate scale factors. ($\rho_s\approx 0.5T_N$.)
This implies that the 2D spin correlations and the 3D magnetic LRO are characterized by a common energy scale that is renormalized by magnetic field. 
It is plausible that the magnetic field changes a balance of competing interactions to make the system more frustrated and suppress magnetic LRO, 
which may appear as a decrease of the characteristic energy and the spin stiffness constant, leading to the suppression of $\xi$ and $1/T_1$. 
In fact, frustration reduces the spin stiffness constant in 2D quantum HAFs \cite{Einarsson95,Bishop15}. 
It is, however, puzzling that the macroscopic susceptibility scarcely depends on the magnetic field in the PM phase. 

$1/T_1$ deviate from the RC scaling relations Eqs.~\eqref{eq:RCC} and \eqref{eq:RCNC} close to $T_N$, signaling an onset of 3D critical slowing down. One may expect a power-law dependence of $1/T_1$ on the reduced temperature $\epsilon=T/T_N-1$ in the critical region, the exponent of which provides information on the universality class of the phase transition. 
We do not pursuit this subject because $^{23}$Na $1/T_1$ is strongly modified by applying field in the critical region, which prevents us from extracting a reliable value of the critical exponent.

\section{\label{sec:discussion}Discussion}

\subsection{\label{sec:HdepT1}Low-energy spin dynamics in the antiferromagnetic phase}

One of the most important characteristics of the spin dynamics in Na$_2$Co$_2$TeO$_6$ is the existence of an intermediate temperature region in the AFM phase (Region III) in which Co spins exhibit unconventional dynamics. 
The magnetic excitation spectrum in Region III ($T_N/2\lesssim T< T_N$) comprises of a broad continuum centered at the ordering wave vector $\mathbf{Q}=(\frac{1}{2},0,0)$ 
rather than a distinct spin-wave mode \cite{Chen21}. 
At low energies probed via $1/T_1$, there remains an unusually large spectral weight of spin fluctuations enhanced strongly with magnetic field. 
In contrast, the spin dynamics at low $T$ in Region IV ($T\lesssim T_N/2$) looks more conventional. 
The continuum is replaced with a gapped spin-wave mode, and $1/T_1$ exhibits a rapid decrease on cooling which evidences the disappearance of low-energy spectral weight. 
The key ingredients in understanding the spin dynamics of Na$_2$Co$_2$TeO$_6$ would thus be 
(i) the origin of a continuum and the large spectral weight at low energies in Region III, (ii) the origin of field enhancement of the low-energy spectral weight, and (iii) the trigger for the formation of a spin-wave mode in Region IV. 

The most likely cause of the broad continuum accompanying a large spectral weight at low energies would be strong damping of spin waves. The appearance of long-lived spin waves at low $T$ is then understood as resulting from diminished damping. 
It is apparent that conventional mechanisms for spin-wave damping in collinear antiferromagnets do not apply because they yield too small damping to account for a broad feature of the excitation spectrum in Region III \cite{Harris71}. 
If allowed by symmetry, cubic anharmonicities in the magnon effective Hamiltonian enable the coupling between one- and two-magnon states which possibly leads to strong damping of the one-magnon mode degenerate with a two-magnon continuum \cite{Zhitomirsky13,Winter17}. 
The cubic terms exist in noncollinear antiferromagnets as well as the Kitaev magnets with off-diagonal interactions and may lead to severe damping at relatively high energies, leaving a low-energy one-magnon mode almost untouched. This is not the case in Na$_2$Co$_2$TeO$_6$ because there is no distinct excitation mode at low energies in Region III on the one hand, and the damping is not so severe at high energies in Region IV on the other \cite{Chen21}. 
Among other things, the fact that the damping changes rapidly its character around a certain temperature ($\sim T_N/2$) seems difficult to be accounted for by a known mechanism for damping which usually gives a smooth variation of magnon lifetime with $T$. 

Although a prime mechanism for the spin-wave damping is unidentified, 
temperature and field variation of the damping may be argued qualitatively based on the general expression of nuclear spin-lattice relaxation rate \cite{Moriya-I},   
\begin{align}\label{eq:T1Sqw}
	\frac{1}{T_1T}=2k_\mathrm{B}\sum_\mathbf{q}\abs{A(\mathbf{q})}^2\frac{\chi^{\prime\prime}(\mathbf{q},\omega_0)}{\omega_0}. 
\end{align}
Here $\chi^{\prime\prime}(\mathbf{q},\omega)$ is the imaginary part of the dynamical spin susceptibility, $\omega_0$ is the nuclear Larmor frequency, and $A(\mathbf{q})$ is the hyperfine form factor determined by a geometry of the nuclear site. Equation~\eqref{eq:T1Sqw} tells us that $1/T_1$ is determined by a spectral weight of spin fluctuations 
at a very low frequency $\omega_0\sim 10^8\mathrm{~s^{-1}}$ ($\hbar\omega_0\sim 0.1\mathrm{~\mu eV}$). 
Since there appears a staggered hyperfine field at the Na sites below $T_N$ as revealed from the NMR line broadening, $A(\mathbf{Q})$ at the ordering wave vector $\mathbf{Q}$ is nonzero and a dominant contribution to $1/T_1T$ will come from $\mathbf{q}\sim\mathbf{Q}$ in the AFM phase. 

The most intuitive view of the results not relying on the specific model is to interpret $1/T_1T$ as a measure of low-energy spin excitations represented by $\chi^{\prime\prime}(\mathbf{q},\omega_0)$. 
It is thus apparent from Figs.~\ref{fig:T1}(b) and \ref{fig:phase} that the low-energy spin excitations remain enhanced in Region III in the AFM phase, especially at high fields of 6$-$7~T as pointed out in the preceding section. 
It is also obvious that the active excitation channels survive to lower temperatures at higher fields. 
The excitation channels activated in Region III almost disappear on entering Region IV as evidenced by a steep decrease of $1/T_1T$ below $\sim$10~K. 

If we take a model of damped harmonic oscillator for spin-wave excitations \cite{Chaikin}, 
we may obtain semi-quantitative information on the spin-wave damping. 
According to the model, we have a contribution of the dynamical susceptibility to $1/T_1T$ in the limit of $\omega_0\rightarrow 0$ 
as $\chi^{\prime\prime}(\mathbf{q},\omega_0)/\omega_0\propto \gamma_\mathbf{q}/\omega_\mathbf{q}^2$ where $\omega_\mathbf{q}$ is an undamped spin-wave frequency and $\gamma_\mathbf{q}$ is a damping constant. 
$\omega_\mathbf{q}$'s are expected to depend only weakly on $T$ except in the vicinity of $T_N$ and unless there is a drastic change of the magnetic structure. The $T$ dependence of $1/T_1T$ at a constant $B$ is thus dominated by that of $\gamma_\mathbf{q}$. 
As to the $B$ dependence, both $\omega_\mathbf{q}$ and $\gamma_\mathbf{q}$ may vary with $B$ and affect the behavior of $1/T_1T$ because of unknown effects of magnetic field on them.

As shown in Figs.~\ref{fig:T1}(b) and \ref{fig:phase}, $1/T_1T$ is peaked around 6$-$7~T at a constant $T$ in Region III. This suggests an increase of $\gamma_\mathbf{q}$ and/or a decrease of $\omega_\mathbf{q}$ with magnetic field. Notice that the latter matches the field-induced reduction of the characteristic energy inferred from the RC analysis of $1/T_1$ in the PM phase. 
At a constant $B$, on the other hand, the $T$ dependence of $1/T_1T$ should be ascribed to that of $\gamma_\mathbf{q}$ as mentioned above. 
Therefore, a broad hump of $1/T_1T$ suggests enhancement of the spin-wave damping at high fields toward the boundary between Regions III and IV. The origin of such strong damping and its unusual $T$ and $B$ dependence is unclear. 
Frustration might play an important role in making the spin excitations incoherent as observed in the triangular-lattice antiferromagnet NaCrO$_2$ below the spin-freezing temperature $T_c\approx 41$~K \cite{Hsieh08}. 
Note, however, that the excitation spectrum of NaCrO$_2$ becomes dispersive below about $0.75T_c$ triggered by the onset of short-range 3D spin correlations, whereas in Na$_2$Co$_2$TeO$_6$ both in-plane and out-of-plane spin correlations show no appreciable change across the boundary between Regions III and IV \cite{Bera17}. 

The presence of magnetic scattering centers is another possibility of strong spin-wave damping. This may provide a reasonable account for the broad continuum in Region III that changes to the dispersive mode in Region IV as described below. 
Structural disorder in Na layers and a related distribution of the interlayer magnetic coupling seem less important because they would give $T$-independent damping. 
It is worth mentioning here that the qualitative change of the excitation spectrum with $T$ in Na$_2$Co$_2$TeO$_6$ resembles the behavior observed in some geometrically frustrated magnets such as the kagom\'e staircase Ni$_3$V$_2$O$_8$ \cite{Ehlers15} and the pyrochlore Gd$_2$Ti$_2$O$_7$ \cite{Paddison21}. These compounds have an intermediate temperature phase just below $T_N$ in which the spin excitations are quasielastic or show only broad features and a low temperature phase with collective excitations. 
The intermediate phases are identified as a partially-disordered state where long-range ordered moments coexist with disordered (paramagnetic) moments, whereas all the sites are ordered in the low temperature phase. This suggests a vital role of partially-disordered moments in the spin-wave damping in these materials. 
Such a scenario may be realized in Na$_2$Co$_2$TeO$_6$ if the AFM phase has triple-$\mathbf{q}$ rather than single-$\mathbf{q}$ zigzag order \cite{Chen21}. 

In the triple-$\mathbf{q}$ ordered state shown in Fig.~\ref{fig:ZigzagTripleQ}(b), three quarters of the Co atoms show noncollinear spin arrangement with a vortex-like texture in the honeycomb planes, while the remaining Co atoms become ``spinless'', which means that the ordered moment is absent, or have only an out-of-plane N\'eel component. The mean fields originating from the in-plane component of the majority spins cancel at the minority site, which may allow the minority spin to fluctuate 
in a large amplitude. The absence of the signal wipeout implies that if the minority spins are not ordered in Region III, they fluctuate in a time scale much faster than the NMR time scale like a paramagnetic moment. 
Spin waves propagating on the majority sites would be strongly damped to give a broad continuum if the resulting quasielastic mode of the minority spins overlaps energetically with the spin-wave mode. 

Since the minority spins are coupled via the third-neighbor interactions 
and possibly via the effective interactions mediated by the majority spins due to quantum fluctuations around the order \cite{Gonzalez19,Seifert19}, 
they would participate in magnetic LRO at low enough temperatures exhibiting slowing of spin fluctuations. 
A broad hump of the magnetic specific heat observed around 10~K \cite{Yao20,Lin21} might be associated with such a change of the minority spin state. 
The spin-wave damping would be diminished as a spectral weight of the quasielastic mode shifts to lower energies, restoring collective excitations at low $T$. 

The low-energy spectral weight of the majority spins arises from the spin-wave damping and is reduced to give a decreasing contribution to $1/T_1T$ at low $T$. On the other hand, slowing of the minority spin fluctuations would contribute a peak or hump of $1/T_1T$ like the case of a magnetic phase transition. The $T$ and $B$ dependence of $1/T_1T$ should be determined by a balance between the two contributions. 
Actually, a feeble anomaly of $1/T_1T$ around 15~K at low fields implies the crossover nature of slowing down rather than a sharp transition with critical dynamics as noted in the previous section. 
A gradual increase of the NMR line splitting below $\sim$15~K \cite{Lee21} might be related to this crossover and the resulting appearance of a static moment on the minority site. 
The broad hump of $1/T_1T$ around 10~K at high fields should then be ascribed primary to the majority spins suffering strong damping even at that temperature for unknown reasons, 
although it might be possible that the minority spin dynamics becomes more critical at higher fields to contribute the hump. 
The $T$ and $B$ dependence of $1/T_1T$ around the boundary between Regions III and IV is complex and is not fully understood in terms of the minority spin ordering. Further investigations are needed to clarify the field-dependent spin dynamics in the AFM phase of Na$_2$Co$_2$TeO$_6$.

\subsection{Comparison with Kitaev candidates}

It has recently been argued whether Na$_2$Co$_2$TeO$_6$ serves as a canonical example of the Kitaev magnet. 
As described in the preceding sections, the spin dynamics in Na$_2$Co$_2$TeO$_6$ displays distinct features from those in other Kitaev candidates such as $\alpha$-RuCl$_3$ and Na$_2$IrO$_3$. 
The magnetic excitation spectrum in the AFM phase of Na$_2$Co$_2$TeO$_6$ is characterized by the sole existence of a broad continuum or a distinct spin-wave mode, both of which has a dominant intensity around the M points of the 2D Brillouin zone ($\mathbf{Q}=(\frac{1}{2},0,0)$ and the equivalents) \cite{Chen21}.
On the other hand, a continuum in $\alpha$-RuCl$_3$ and Na$_2$IrO$_3$ is centered at the $\Gamma$ point (zone center) and coexists with spin-wave modes below $T_N$ \cite{Banerjee17,KimJ20}. 
The major excitations around the M point in Na$_2$Co$_2$TeO$_6$ are possibly ascribed to large third-neighbor coupling $J_3$ suggested from the powder INS \cite{Samarakoon21,Lin21,KimC22,Sanders22}. This is known to stabilize zigzag order but to counteract the formation of a Kitaev QSL \cite{Liu20,Liu21}. 
In the context of Kitaev physics, the honeycomb cobaltate Na$_3$Co$_2$SbO$_6$ seems more promising because it exhibits intense excitations around the $\Gamma$ point probably due to smaller $J_3$ \cite{Songvilay20,KimC22}. 

On the field evolution of the low-energy spin dynamics, it is interesting to compare our results of $1/T_1$ with those for other Kitaev candidates exhibiting a similar field-induced transition from an AFM phase to a spin-disordered phase. 
To the best of the authors' knowledge, $\alpha$-RuCl$_3$ is only one such example with extensive field-dependent NMR studies \cite{Baek17,Zheng17,Jansa18,Nagai20}. 
The AFM phase of another well-studied candidate Na$_2$IrO$_3$ is robust against a field \cite{Ye12}. 
In fact, $1/T_1$ at the Na sites in Na$_2$IrO$_3$ is insensitive to field and shows conventional behaviors above and below $T_N$ \cite{Takahashi19}. 

$\alpha$-RuCl$_3$ seems to be the best reference as it shows zigzag order like Na$_2$Co$_2$TeO$_6$. The in-plane critical field $B_c$ to the high-field disordered phase is 7$-$8~T \cite{Johnson15,Sears17,Wolter17}. 
Most of the NMR experiments on $\alpha$-RuCl$_3$, however, focused on the behavior near and above $B_c$ 
and the field evolution of $1/T_1$ has not been investigated systematically in the low-field region. 
Although a direct comparison of the results is limited to a narrow field range, the field response of $1/T_1$ has distinct differences between the two compounds. This may reflect the presence of the intermediate temperature region (Region III) characterized by a substantial low-energy spectral weight in Na$_2$Co$_2$TeO$_6$ 
and the excitation continuum coexisting with spin-wave modes in $\alpha$-RuCl$_3$. 
Indeed, $1/T_1$ at the $^{35}$Cl site in the AFM phase of $\alpha$-RuCl$_3$ is relatively insensitive to field except above the field $B_c^\prime=7.1$~T $(<B_c)$ where gapless magnon excitations have been suggested \cite{Nagai20}. 
This shows a marked contrast to strong field enhancement of $^{23}$Na $1/T_1$ in Region III of Na$_2$Co$_2$TeO$_6$ starting far below the critical field $B_c\approx 10$~T. 
The field-insensitive response of $1/T_1$ at $B<B_c^\prime$ in $\alpha$-RuCl$_3$ would be due to gapped magnon excitations, 
but at $T<4$~K well below $T_N=6.5$~K, $1/T_1$ is contributed by a residual mode that grows with field on crossing $B_c^\prime$ and becomes dominant above $B_c$ \cite{Baek17,Nagai20}. 
Such a mode was not detected in Na$_2$Co$_2$TeO$_6$ down to 3.5~K and may be associated with the continuum around the $\Gamma$ point in $\alpha$-RuCl$_3$ identified as excitations inherent to Kitaev QSLs.  

Despite the apparent differences in the low-energy spin dynamics probed by $1/T_1$, the two compounds have a lot of similarities in their response to magnetic field; closing of the AFM phase suggesting the existence of a quantum critical point, the possible appearance of a field-induced QSL phase, and so on. 
It is often encountered in strongly frustrated magnets that the magnetic LRO is controlled by sub-leading interactions instead of the leading one like the Kitaev coupling and by external perturbations such as magnetic field, pressure, and in some cases spin defects.  
Since the low-energy sector of a magnetic excitation spectrum is very much affected and reconstructed by these interactions and perturbations, there will be a wide variety of phases and behaviors in real materials which at first glance look very different. 
Concerning the present case, it might be possible that the unusual temperature and field evolution of low-energy spin dynamics in Na$_2$Co$_2$TeO$_6$ is described in terms of a generalized Kitaev model by including necessary factors. 
It is worth noting that for the reported values of the Kitaev coupling $K$ \cite{Songvilay20,KimC22,Samarakoon21,Sanders22}, Region III is around or lower than the crossover temperature $T_H\sim 0.375K$ below which localized $Z_2$ fluxes and itinerant Majorana fermions are expected to emerge \cite{Yoshitake16}.  
The close resemblance between $1/T_1T$ and $C_m/T$, however, implies confinement of the fractionalized particles to magnons. 
Our findings on the low-energy spin dynamics of Na$_2$Co$_2$TeO$_6$ will thus provide new insights into Kitaev-derived spin models as well as more conventional models on the honeycomb lattice, promoting future studies in this research field. 
From an experimental side, field-dependent microscopic measurements complementing NMR, such as neutron, Raman, and terahertz spectroscopies, are highly required.

\section{Summary and Conclusions}

We have measured $^{23}$Na NMR in the honeycomb lattice antiferromagnet Na$_2$Co$_2$TeO$_6$ to elucidate the phases and the underlying low-energy spin dynamics in a wide range of temperature and magnetic field. 
The magnetic phase diagram was constructed using the microscopic quantities measured by NMR. The persistence of AFM order up to a field of 9~T was confirmed from the magnetic shift and broadening of the NMR spectrum and the rapidly decreasing $1/T_1$ at low $T$. 

The AFM phase is divided into two distinct temperature regions that exhibit contrasting low-energy dynamics and its field response. In the intermediate temperature region just below $T_N$ (Region III), 
there exists an appreciable low-energy spectral weight of spin fluctuations that contributes to $1/T_1$ and is enhanced strongly with magnetic field. 
The low temperature region below $\sim T_N/2$ (Region IV) is characterized by a loss of this low-energy weight as evidenced via a rapid decrease of $1/T_1$ which is less field-dependent. 
The qualitative change of the low-energy spin dynamics across the boundary between the two regions is consistent with the fact that the magnetic excitation spectrum at higher energies displays an incoherent feature in Region III and a gapped dispersive mode in Region IV \cite{Chen21}. 

We interpreted the lack of a dispersive mode and the presence of a significant low-energy spectral weight in Region III as arising from strong spin-wave damping. The appearance of a dispersive mode in Region IV is then ascribed to weakening of the damping. 
As a possible scenario, 
we suggested a partially-disordered state in Region III with the triple-$\mathbf{q}$ magnetic structure formed by superposing three equivalent zigzag patterns. 
In this scenario, the partially-disordered moment experiencing a vanishing mean field acts as a strong scatterer of spin waves propagating on the ordered sites in Region III   
and acquires an ordered moment to take part in the collective excitations in Region IV. 
The scattering hence weakens to restore spin-wave excitations at sufficiently low temperatures. 
The scenario, however, cannot fully account for the complex behavior of $1/T_1T$ around the boundary between Regions III and IV and needs further investigation.  

We also identified a temperature region with field-dependent 2D spin correlations in the PM phase near $T_N$ (Region II). The $T$ dependence of $1/T_1$ in Region II is well reproduced using the renormalized-classical scaling form for 2D quantum antiferromagnets. 
The field suppression of $1/T_1$ due to a reduction of the in-plane correlation length is described by a monotonic decrease of the spin stiffness constant with magnetic field, suggesting the existence of a high-field disordered phase in the limit of vanishing spin stiffness. 
The fact that the spin stiffness constant scales with $T_N$ as a function of magnetic field implies a common energy scale for the 2D spin correlations and 3D magnetic LRO. 
The magnetic phases and the spin dynamics may be controlled by field-tuning this energy scale, which is likely caused by cancellation of frustrating interactions including an effect of external magnetic fields.

\begin{acknowledgments}
We thank Y. Itoh and H. Kusunose for valuable discussions. This work was partially supported by Grants-in-Aid for Scientific Research, MEXT, Japan (KAKENHI Grant No.15K05149).
\end{acknowledgments}

\appendix*
\section{Inverse Laplace transform analysis of $1/T_1$ in the antiferromagnetic phase}
It is well known that the stretched exponential analysis of magnetization recovery dictates a specific form of the distribution function for $1/T_1$ \cite{Johnston05,Johnston06} 
which may not reflect the true distribution of $1/T_1$. 
One should thus be careful in making concrete statements about results of $1/T_1$ 
when there exist inequivalent nuclear sites and/or some domains exhibiting distinct spin-lattice relaxation. 
In this Appendix, we present analysis of magnetization recovery in the AFM phase where the stretching exponent $\beta$ as well as $1/T_1$ is strongly $T$ dependent, based on the method of so-called inverse Laplace transform (ILT) which can deduce the probability distribution function $P(1/T_1)$, i.e., the histogram of $1/T_1$. This method has recently been applied successfully to analyze spatially-inhomogeneous spin-lattice relaxation in high-$T_c$ cuprates \cite{Singer20,Arsenault20}. 

The ILT analysis assumes that each nucleus decays as a linear combination of normal modes with a definite relaxation rate, but the rate is heterogeneous over the sample and is described by a distribution function $P(1/T_1)$. For nuclei with $I=3/2$, the magnetization recovery $M(t)$ may be expressed in a discrete form for $P(1/T_1)$ as 
\begin{align}
	M(t) = \sum_{j=1}^N \Big[1 - A\sum_{k=1}^3 \alpha_k e^{-\lambda_k t/T_{1j}}\Big] &P(1/T_{1j}).
\end{align}
Here $N$ is the number of bins for $P(1/T_1)$, $A$ is a degree of inversion, $\{\lambda_k\}=\{1,3,6\}$ are mode eigenvalues, and $\{\alpha_k\}$ are amplitudes of the corresponding modes satisfying $\sum_k \alpha_k=1$. 
The summation $\sum_j P(1/T_{1j}) = M(\infty)$ is the equilibrium magnetization. 
The ILT analysis deduces $\{P(1/T_{1j})\}$ numerically from recovery data $\{M(t_i)\}$ ($t_i$ being the delay time) without assuming any functional form of $P(1/T_1)$. 
For technical details, see the Supplemental Material of Ref.~\onlinecite{Singer20} and references therein. 
We take 250 bins for $P(1/T_1)$ equally spaced on a logarithmic scale ranging from $10^{-2}\mathrm{~s^{-1}}\leq 1/T_{1j}\leq 10^6\mathrm{~s^{-1}}$. Tikhonov regularization method was employed to find the optimal solution. The resulting probability distribution is then normalized as $\sum_j P(1/T_{1j})\Delta_P=1$ where $\Delta_P$ is the logarithmic bin spacing. 

When using polycrystals, one cannot determine $\{\alpha_k\}$ uniquely because the initial ($t=0$) populations of the nuclear level are not known exactly. 
We examined the following models for $\{\alpha_k\}$ to perform ILT utilizing results of the fitting of $\{M(t_i)\}$ to Eq.~\eqref{eq:recovery}: 
\textit{Model A}, $\{\alpha_k\}$ taken as those determined at each temperature; 
\textit{Model B}, $\{\alpha_k\}$ fixed to the values at 30~K just above $T_N$; and 
\textit{Model C}, $\{\alpha_k\}$ taken as the average values in Region III ($T_N/2\lesssim T<T_N$) where $\{\alpha_k\}$ are almost $T$ independent. 
We found that $P(1/T_1)$ is relatively insensitive to the choice of the above models for $\{\alpha_k\}$. 
\begin{figure}[t]
\includegraphics[width=7.7cm]{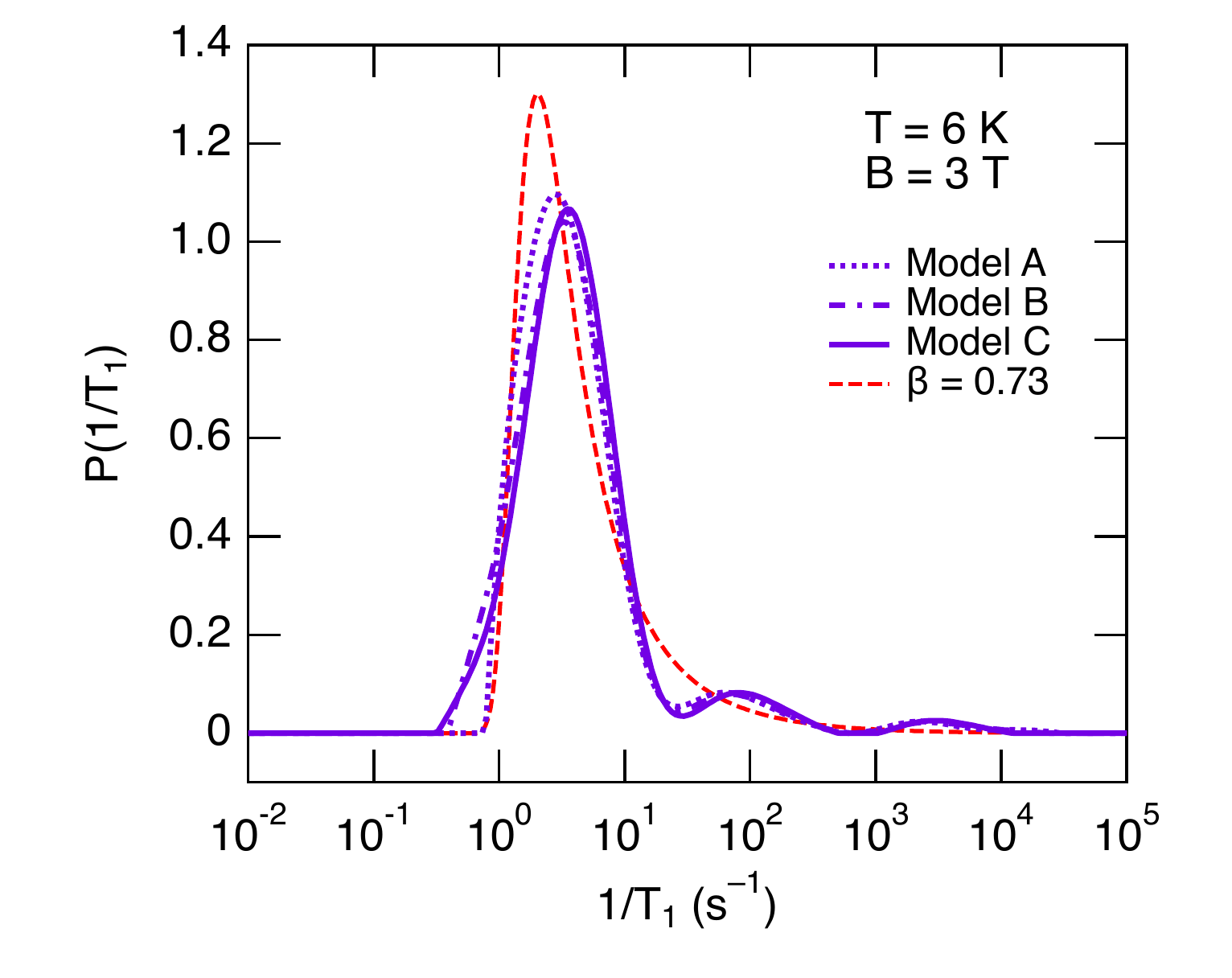}
\caption{\label{fig:P3T6K}Probability distribution function $P(1/T_1)$ deduced from the ILT analysis of the magnetization recovery at $T=6$~K at $B=3$~T. Mode amplitudes $\{\alpha_k\}$ of each model are as follows; Model A (dotted line), $\{0.24,0,0.76\}$; Model B (dashed-dotted line), $\{0.17,0.12,0.71\}$; Model C (solid line), $\{0.17,0.22,0.61\}$. The dashed line is $P(1/T_1)$ for the stretched exponential recovery with $\beta=0.73$ and $1/T_1=3.4\mathrm{~s^{-1}}$ calculated using Eq.~(11) of Ref.~\onlinecite{Johnston06}.} 
\end{figure}

Figure~\ref{fig:P3T6K} displays the distribution functions obtained from the recovery at $T=6$~K at $B=3$~T shown in Fig.~\ref{fig:T1}(d). 
The $P(1/T_1)$'s are almost identical, peaked at $1/T_1\sim 3\mathrm{~s^{-1}}$ and having nearly a decade width, except that $P(1/T_1)$ for Model A exhibits sharper cutoff at the side of low relaxation rates than Models B and C. Wiggly subpeaks appearing at the side of high relaxation rates would be an oscillatory artifact \cite{Choi21}. In fact, a position of the subpeak depends on the details of the ILT analysis such as the number of bins and the choice of the regularization (smoothing) factor. 
The ILT fits to $\{M(t_i)\}$ (not shown) were as good as the stretched exponential fit. 

The $P(1/T_1)$'s deduced from the ILT analysis resemble the one for the stretched exponential function in that they are single-peaked and have a long tail at the side of high relaxation rates \cite{Johnston05,Johnston06}. 
For comparison, we calculated $P(1/T_1)$ numerically for the stretched exponential function using the expression given in Ref.~\onlinecite{Johnston06}. The exponent $\beta=0.73$ and $1/T_1=3.4\mathrm{~s^{-1}}$ obtained from the fitting to Eq.~\eqref{eq:recovery} were used. 
The overall line shape of the ILT $P(1/T_1)$'s is well reproduced by $P(1/T_1)$ for the stretched exponential function. 
\begin{figure}[t]
\includegraphics[width=8.2cm]{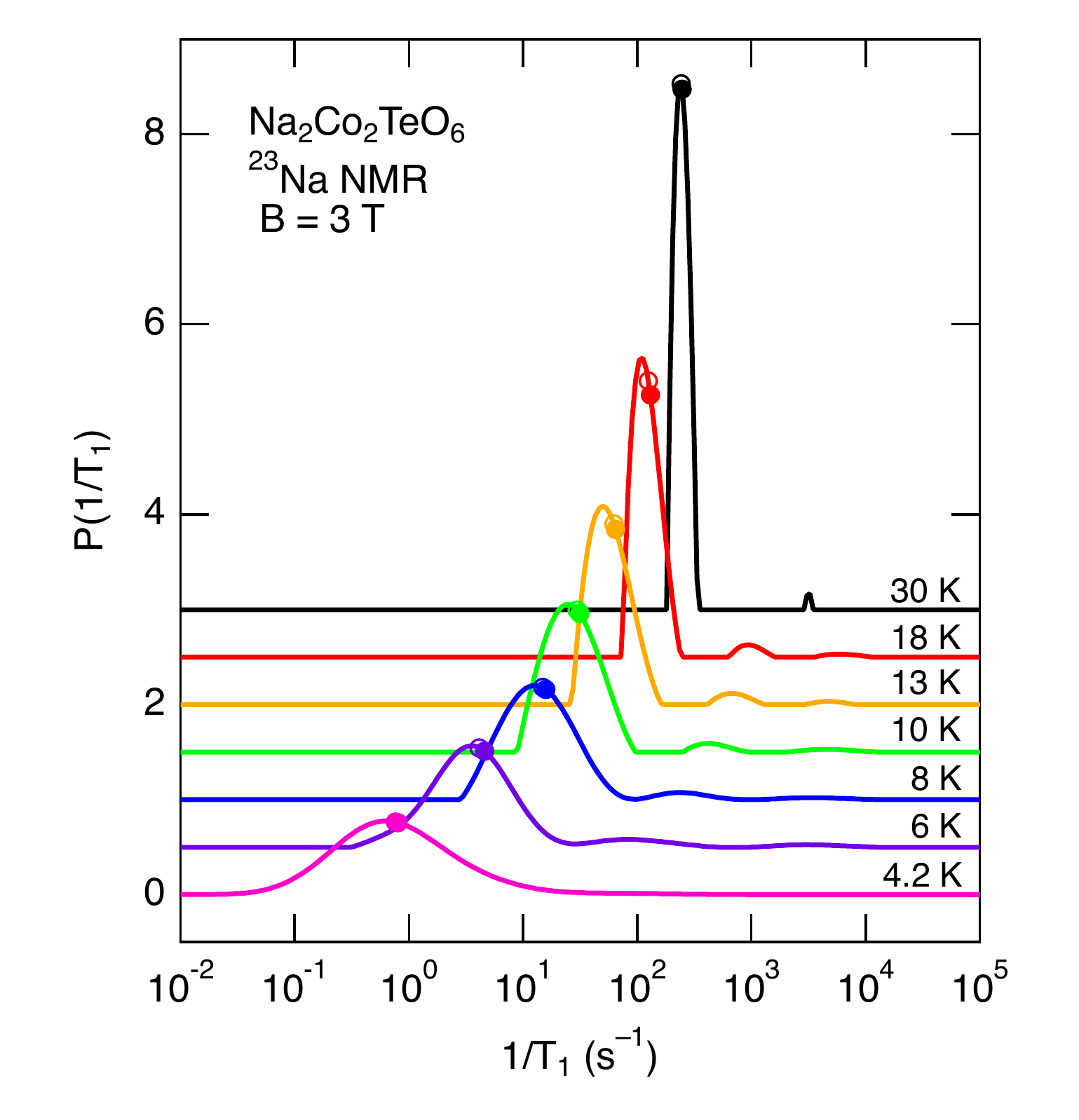}
\caption{\label{fig:P3T}Temperature evolution of the probability distribution function $P(1/T_1)$ below $T=30$~K at $B=3$~T. Filled bullets represent the location of the log means $1/T_1^\mathrm{lm}$ of $P(1/T_1)$. Open bullets are $1/T_1$ determined by the stretched exponential fitting using Eq.~\eqref{eq:recovery}.}
\end{figure}

Figure~\ref{fig:P3T} shows $T$ evolution of $P(1/T_1)$ below $T=30$~K at $B=3$~T 
\footnote{The mode amplitudes at 30~K above $T_N$ were taken as $\{\alpha_k\}=\{0.17,0.12,0.71\}$ using the result of stretched exponential fitting.}. 
We adopted Model C for $\{\alpha_k\}$ based on the idea that $\{\alpha_k\}$ would not depend strongly on $T$ in the AFM phase where the NMR line width does not vary strongly with $T$. 
As the temperature is lowered, $P(1/T_1)$ becomes progressively broader, exhibiting an oscillatory tail at the side of high relaxation rates. No extra peak showing a distinct $T$ variation of $1/T_1$ from the main peak appears. 
The log means $1/T_1^\mathrm{lm}$ of the distribution function defined by 
$\ln(1/T_1^\mathrm{lm}) = \sum_j \ln(1/T_{1j}) P(1/T_{1j})\Delta_P$ \cite{Singer20} 
are marked as filled bullets on each $P(1/T_1)$ curve in Fig.~\ref{fig:P3T}. 
They are in good agreement with $1/T_1$ determined by the stretched exponential fitting and shown as open bullets. 
The line shape is also consistent with the stretched exponential analysis at each temperature. 
These facts indicate that the distribution of $1/T_1$ at the Na sites and its $T$ evolution are well captured by the phenomenological stretched exponential analysis, 
justifying the resulting $1/T_1$ and $\beta$ as representing the average relaxation rate and the distribution of $1/T_1$ in the AFM phase.

We performed the same analysis at $B=7$~T. The results are qualitatively similar to those at $B=3$~T. This means that the low-energy spin dynamics does not change much with field in a range covered in the present study as far as the distribution of $1/T_1$ is concerned.

\bibliography{jkiku_NCTO_NMR_ref.bbl}

\end{document}